\documentclass[sigconf]{acmart}

\usepackage{balance}

\def\BibTeX{{\rm B\kern-.05em{\sc i\kern-.025em b}\kern-.08emT\kern-.1667em\lower.7ex\hbox{E}\kern-.125emX}}

\usepackage{lipsum,adjustbox,xcolor}
\usepackage{setspace}
\usepackage{booktabs, tabularx, colortbl}
\usepackage{subfigure}
\usepackage{array, xspace, url, amsmath, bm, array,amsfonts, balance,multirow, color, enumitem,kantlipsum, float, multicol,ragged2e, soul}

\usepackage[nounderscore]{syntax}
\usepackage{newfloat}
\usepackage{mdframed}

\DeclareFloatingEnvironment[
  fileext   = logr,
  listname  = {List of Grammars},
  name      = Grammar,
  placement = !htbp
]{BNF}

\usepackage[capitalize,nameinlink]{cleveref}
\hypersetup{%
	bookmarksnumbered, bookmarksopen=true, bookmarksopenlevel=1,%
}
\crefname{figure}{Figure}{Figures}
\crefname{listing}{Query}{Queries}
\crefname{section}{\S}{\S}
\crefname{table}{Table}{Tables}
\crefname{BNF}{Grammar}{Grammars}
\crefname{algorithm}{Algorithm}{Algorithms}

\definecolor{darkblue}{rgb}{0,0,0.5}
\definecolor{darkred}{rgb}{0.5,0,0}
\definecolor{darkgreen}{rgb}{0,0.5,0}
\definecolor{balanceblue}{rgb}{0.25,0,0.25}
\definecolor{balance}{rgb}{0.25,0.25,1.0}
\definecolor{extensionA}{rgb}{0.25,0.5,1.0}
\definecolor{extensionB}{rgb}{1.0,0.25,0.25}
\definecolor{extensionC}{rgb}{0,0.5,0.5}
\definecolor{extensionD}{RGB}{75,0,130}
\definecolor{extensionE}{rgb}{0.1,0.1,0.5}
\definecolor{tablecol}{rgb}{0.9,0.95,0.95}
\definecolor{comment}{RGB}{105,105,105}
\hypersetup{
	colorlinks=true,
	urlcolor=darkblue,
	linkcolor=darkblue,
	citecolor=darkblue
}

\DeclareRobustCommand{\hlgray}[1]{{\sethlcolor{gray!30}\hl{#1}}}

\makeatletter
\newdimen\SOUL@dimen 
\def\SOUL@ulunderline#1{{%
    \setbox\z@\hbox{#1}%
    \SOUL@dimen=\wd\z@ 
    \dimen@i=\SOUL@uloverlap
    \advance\SOUL@dimen2\dimen@i 
    \rlap{%
        \null
        \kern-\dimen@i
        \SOUL@ulcolor{\SOUL@ulleaders\hskip\SOUL@dimen}
    }%
    \unhcopy\z@
}}
\makeatother

\usepackage{scalerel,stackengine}
\stackMath
\newcommand\reallywidehat[1]{%
	\savestack{\tmpbox}{\stretchto{%
			\scaleto{%
				\scalerel*[\widthof{\ensuremath{#1}}]{\kern.1pt\mathchar"0362\kern.1pt}%
				{\rule{0ex}{\textheight}}
			}{\textheight}%
		}{2.25ex}}%
	\stackon[-6.9pt]{#1}{\tmpbox}%
}

\newcommand{\ie}{\emph{i.e.,}\xspace}
\newcommand{\eg}{\emph{e.g.,}\xspace}
\newcommand{\cf}{\emph{c.f.,}\xspace}
\newcommand{\sys}{\textsf{HyperService}\xspace}
\newcommand{\HSL}{\textsf{HSL}\xspace}
\newcommand{\USM}{\textsf{USM}\xspace}
\newcommand{\BIP}{\textsf{UIP}\xspace}
\newcommand{\UIP}{\textsf{UIP}\xspace}

\newcommand{\ISC}{\textsf{ISC}\xspace}
\newcommand{\ISCs}{\textsf{ISC}s\xspace}
\newcommand{\VEN}{\textsf{VES}\xspace}
\newcommand{\NSB}{\textsf{NSB}\xspace}

\newcommand{\VENs}{\textsf{VESes}\xspace}
\newcommand{\dApp}{\textsf{dApp}\xspace}
\newcommand{\CLI}{\textsf{CLI}\xspace}
\newcommand{\dApps}{\textsf{dApps}\xspace}

\newcommand{\gt}{$\mathcal{G}_T$\xspace}
\newcommand{\rgt}{\mathcal{G}_T}
\newcommand{\cert}{\textcolor{balance}{\textsf{Cert}}\xspace}

\newcommand{\haccount}{\textcolor{darkblue}{\emph{\textsf{account}}}\xspace}
\newcommand{\hcontract}{\textcolor{darkblue}{\emph{\textsf{contract}}}\xspace}
\newcommand{\hpay}{\textcolor{darkblue}{\emph{\textsf{payment}}}\xspace}
\newcommand{\hinvo}{\textcolor{darkblue}{\emph{\textsf{invocation}}}\xspace}
\newcommand{\hprecon}{\textcolor{darkblue}{\emph{\textsf{precondition}}}\xspace}
\newcommand{\hddl}{\textcolor{darkblue}{\emph{\textsf{deadline}}}\xspace}
\newcommand{\hxcoin}{\textcolor{extensionA}{\emph{\textsf{xcoin}}}\xspace}
\newcommand{\hycoin}{\textcolor{extensionA}{\emph{\textsf{ycoin}}}\xspace}
\newcommand{\hzcoin}{\textcolor{extensionA}{\emph{\textsf{zcoin}}}\xspace}
\newcommand{\hncoin}{\textcolor{extensionA}{\emph{\textsf{ncoin}}}\xspace}
\newcommand{\hop}{\textcolor{darkblue}{\emph{\textsf{op}}}\xspace}
\newcommand{\husing}{\textcolor{darkblue}{\emph{\textsf{using}}}\xspace}
\newcommand{\hfrom}{\textcolor{darkblue}{\emph{\textsf{from}}}\xspace}
\newcommand{\hto}{\textcolor{darkblue}{\emph{\textsf{to}}}\xspace}
\newcommand{\has}{\textcolor{darkblue}{\emph{\textsf{as}}}\xspace}
\newcommand{\hwith}{\textcolor{darkblue}{\emph{\textsf{with}}}\xspace}
\newcommand{\hbefore}{\textcolor{darkblue}{\emph{\textsf{before}}}\xspace}
\newcommand{\hafter}{\textcolor{darkblue}{\emph{\textsf{after}}}\xspace}
\newcommand{\himport}{\textcolor{darkblue}{\emph{\textsf{import}}}\xspace}
\newcommand{\hchainx}{\textcolor{balanceblue}{\emph{\textsf{ChainX}}}\xspace}
\newcommand{\hchainy}{\textcolor{balanceblue}{\emph{\textsf{ChainY}}}\xspace}
\newcommand{\hchainz}{\textcolor{balanceblue}{\emph{\textsf{ChainZ}}}\xspace}

\newcommand{\unknown}{\textsf{unknown}\xspace}
\newcommand{\init}{\textsf{init}\xspace}
\newcommand{\initd}{\textsf{inited}\xspace}
\newcommand{\id}{\textsf{id}\xspace}
\newcommand{\od}{\textsf{od}\xspace}

\newcommand{\open}{\textsf{open}\xspace}
\newcommand{\opend}{\textsf{opened}\xspace}
\newcommand{\close}{\textsf{closed}\xspace}

\newcommand{\cort}{\textsf{correct}\xspace}

\newcommand{\mt}{\mathcal{T}\xspace}

\newcommand{\md}{\mathcal{D}\xspace}
\newcommand{\mv}{\mathcal{V}\xspace}
\newcommand{\mx}{\mathcal{X}\xspace}
\newcommand{\ms}{\mathcal{S}\xspace}
\newcommand{\mathp}{\mathcal{P}\xspace}
\newcommand{\me}{\mathcal{E}\xspace}
\newcommand{\ma}{\mathcal{A}\xspace}

\newcommand{\mk}{\mathcal{K}\xspace}
\newcommand{\mi}{\mathcal{I}\xspace}
\newcommand{\wtt}{\widetilde{T}\xspace}
\newcommand{\bco}{\textbf{:}\xspace}
\newcommand{\sig}{\textsf{Sig}\xspace}
\newcommand{\sid}{\textcolor{darkgreen}{\textsf{sid}}\xspace}
\newcommand{\cid}{\textcolor{darkblue}{\textsf{cid}}\xspace}

\newcommand{\sigd}{\sig_{\sid}^\mathcal{D}\xspace}
\newcommand{\sigv}{\sig_{\sid}^\mathcal{V}\xspace}

\newcommand{\atte}{\textcolor{darkred}{\textsf{Cert}}\xspace}
\newcommand{\realatte}{\textcolor{darkred}{\textsf{Atte}}\xspace}
\newcommand{\merk}{\textcolor{darkred}{\textsf{Merk}}\xspace}
\newcommand{\hamt}{\textcolor{balanceblue}{\textsf{meta}}\xspace}
\newcommand{\amt}{\textcolor{balanceblue}{\textsf{amt}}\xspace}
\newcommand{\dst}{\textcolor{balanceblue}{\textsf{dst}}\xspace}
\newcommand{\Op}{\textcolor{darkblue}{\textsf{Op}}\xspace}

\newcommand{\pbip}{\textcolor{extensionC}{$\textsf{Prot}_{\BIP}$}\xspace}
\newcommand{\hpbip}{\textcolor{extensionC}{$\textsf{H-Prot}_{\BIP}$}\xspace}
\newcommand{\fbip}{\textcolor{extensionC}{$\mathcal{F}_{\BIP}$}\xspace}
\newcommand{\idealbip}{\textcolor{extensionC}{$\mathcal{I}_{\mathcal{F}_{\BIP}}$}\xspace}

\newcommand{\fven}{\textcolor{extensionC}{$\textsf{Prot}_{\VEN}$}\xspace}
\newcommand{\fisc}{\textcolor{extensionC}{$\textsf{Prot}_{\ISC}$}\xspace}
\newcommand{\fdapp}{\textcolor{extensionC}{$\textsf{Prot}_{\CLI}$}\xspace}
\newcommand{\fbc}{\textcolor{extensionC}{$\mathcal{F}_{\textsf{blockchain}}$}\xspace}

\newcommand{\pbc}{\textcolor{extensionC}{$\textsf{Prot}_{\textsf{BC}}$}\xspace}

\newcommand{\fnsb}{\textcolor{extensionC}{$\textsf{Prot}_{\NSB}$}\xspace}

\newcommand{\pven}{\textcolor{extensionC}{$\mathcal{P}_{\VEN}$}\xspace}
\newcommand{\ipven}{\textcolor{extensionC}{$\mathcal{P}_{\VEN}^{\mi}$}\xspace}
\newcommand{\ripven}{\textcolor{extensionC}{\mathcal{P}_{\VEN}^{\mi}}\xspace}
\newcommand{\pdapp}{\textcolor{extensionC}{$\mathcal{P}_{\CLI}$}\xspace}
\newcommand{\ipdapp}{\textcolor{extensionC}{$\mathcal{P}_{\CLI}^{\mi}$}\xspace}
\newcommand{\pisc}{\textcolor{extensionC}{$\mathcal{P}_{\ISC}$}\xspace}
\newcommand{\ipisc}{\textcolor{extensionC}{$\mathcal{P}_{\ISC}^{\mi}$}\xspace}
\newcommand{\pnsb}{\textcolor{extensionC}{$\mathcal{P}_{\NSB}$}\xspace}
\newcommand{\ipnsb}{\textcolor{extensionC}{$\mathcal{P}_{\NSB}^{\mi}$}\xspace}
\newcommand{\pybc}{\textcolor{extensionC}{$\mathcal{P}_{\textsf{BC}}$}\xspace}
\newcommand{\ipybc}{\textcolor{extensionC}{$\mathcal{P}_{\textsf{BC}}^{\mi}$}\xspace}

\newcommand{\timer}{\emph{\textsf{timer}}\xspace}
\newcommand{\public}{\textcolor{extensionE}{\textsf{public}}\xspace}
\newcommand{\private}{\textcolor{extensionE}{\textsf{private}}\xspace}
\newcommand{\ord}{\textcolor{extensionE}{\textsf{override}}\xspace}

\newcommand{\secr}{\textcolor{extensionD}{\textsf{SessionCreate}}}

\newcommand{\tmr}{\textcolor{extensionD}{\textsf{TermExecution}}}
\newcommand{\rto}{\textcolor{extensionD}{\textsf{ReqTransOpen}}}

\newcommand{\rtc}{\textcolor{extensionD}{\textsf{ReqTransClose}}}

\newcommand{\rti}{\textcolor{extensionD}{\textsf{ReqTransInit}}}
\newcommand{\rtid}{\textcolor{extensionD}{\textsf{ReqTransInited}}}
\newcommand{\rtod}{\textcolor{extensionD}{\textsf{ReqTransOpened}}}

\newcommand{\sread}{\textcolor{extensionD}{\textsf{StatusRead}}}
\newcommand{\siga}{\sig_{\sid}^{\mathcal{P}_a}\xspace}
\newcommand{\sigz}{\sig_{\sid}^{\mathcal{P}_z}\xspace}
\newcommand{\sigpt}{\sig_{\sid}^{\mathcal{P}}\xspace}

\newcommand{\pcmpl}{\textcolor{extensionD}{\textsf{PostCompiliation}}}
\newcommand{\xpcmplb}{\textsf{PostCompiliation}\xspace}

\newcommand{\xinittb}{\textsf{InitTrans}\xspace}

\newcommand{\sdited}{\textcolor{extensionD}{\textsf{SInitedTrans}}}
\newcommand{\xsditedb}{\textsf{SInitedTrans}\xspace}

\newcommand{\rvited}{\textcolor{extensionD}{\textsf{RInitedTrans}}}
\newcommand{\xrvitedb}{\textsf{RInitedTrans}\xspace}

\newcommand{\xitdtb}{\textsf{InitedTrans}\xspace}

\newcommand{\opt}{\textcolor{extensionD}{\textsf{OpenTrans}}}
\newcommand{\xoptb}{\textsf{OpenTrans}\xspace}
\newcommand{\opdt}{\textcolor{extensionD}{\textsf{OpenedTrans}}}

\newcommand{\xopdtb}{\textsf{OpenedTrans}\xspace}
\newcommand{\cldt}{\textcolor{extensionD}{\textsf{ClosedTrans}}}

\newcommand{\xcldtb}{\textsf{ClosedTrans}\xspace}

\newcommand{\clt}{\textcolor{extensionD}{\textsf{CloseTrans}}}

\newcommand{\xcltb}{\textsf{CloseTrans}\xspace}

\newcommand{\redem}{\textcolor{extensionD}{\textsf{Redeem}}}
\newcommand{\probe}{\textcolor{extensionD}{\textsf{Watching}}}

\newcommand{\xcrc}{\textcolor{extensionD}{\textsf{CreateContract}}\xspace}

\newcommand{\xcrcb}{\textsf{CreateContract}\xspace}
\newcommand{\xskf}{\textcolor{extensionD}{\textsf{StakeFund}}\xspace}

\newcommand{\xskfb}{\textsf{StakeFund}\xspace}

\newcommand{\insu}{\textcolor{extensionD}{\textsf{InsuranceClaim}}}
\newcommand{\xinsub}{\textsf{InsuranceClaim}\xspace}

\newcommand{\setc}{\textcolor{extensionD}{\textsf{SettleContract}}}
\newcommand{\xsetcb}{\textsf{SettleContract}\xspace}

\newcommand{\revs}{\textsf{revs}\xspace}
\newcommand{\contract}{\emph{\textsf{contract}}\xspace}

\newcommand{\tsopen}{\textsf{ts}_\textsf{open}\xspace}
\newcommand{\tsclose}{\textsf{ts}_\textsf{closed}\xspace}
\newcommand{\stake}{\textsf{stake}\xspace}

\newcommand{\xdct}{\textcolor{extensionD}{\textsf{DiscreteTimer}}\xspace}
\newcommand{\xdctb}{\textsf{DiscreteTimer}\xspace}
\newcommand{\xmkp}{\textcolor{extensionD}{\textsf{MerkleProof}}\xspace}

\newcommand{\xmkpb}{\textsf{MerkleProof}\xspace}
\newcommand{\xexec}{\textcolor{extensionD}{\textsf{Exec}}\xspace}
\newcommand{\exec}{\textcolor{extensionD}{\textsf{Exec}}}

\newcommand{\xadda}{\textcolor{extensionD}{\textsf{AddAction}}\xspace}

\newcommand{\xaddab}{\textsf{AddAction}\xspace}
\newcommand{\xbh}{\textcolor{extensionD}{\textsf{BlockHeight}}\xspace}

\newcommand{\xbhb}{\textsf{BlockHeight}\xspace}

\newcommand{\cc}{\textcolor{extensionD}{\textsf{CloseureClaim}}}
\newcommand{\xccb}{\textsf{CloseureClaim}\xspace}

\newcommand{\cw}{\textcolor{extensionD}{\textsf{CloseureWatching}}}
\newcommand{\xcwb}{\textsf{CloseureWatching}\xspace}

\newcommand{\mb}{\mathcal{B}}
\newcommand{\txp}{\textsf{TxPool}\xspace}
\newcommand{\acp}{\textsf{ActionPool}\xspace}
\newcommand{\stp}{\textsf{StatusPool}\xspace}
\newcommand{\txmt}{\textsf{TxMT}\xspace}
\newcommand{\acmt}{\textsf{ActionMT}\xspace}
\newcommand{\stmt}{\textsf{StatusMT}\xspace}
\newcommand{\statemt}{\textsf{StateMT}\xspace}
\newcommand{\ledger}{\textsf{Ledger}\xspace}

\newcommand{\first}{\textsf{(i)}\xspace}
\newcommand{\second}{\textsf{(ii)}\xspace}
\newcommand{\third}{\textsf{(iii)}\xspace}
\newcommand{\forth}{\textsf{(iv)}\xspace}

\makeatletter
\newif\if@restonecol
\makeatother

\usepackage[linesnumbered,boxed, figure]{algorithm2e}
\usepackage[noend]{algpseudocode}

\SetCommentSty{mycommfont}

\usepackage[skip=3pt]{caption}

\newcommand{\paraspace}{\vspace{0.02in}}
\newcommand{\parab}[1]{\paraspace\noindent{\bf #1}}

\setcopyright{none}
\renewcommand\footnotetextcopyrightpermission[1]{}
\begin{document}
\fancyhead{}
\title{\sys: Interoperability and Programmability \\ Across Heterogeneous Blockchains}
\titlenote{This is an extended version of the material originally published in ACM CCS 2019.}
\author{Zhuotao Liu$^{1,2}$ \enskip Yangxi Xiang$^3$ \enskip Jian Shi$^4$ \enskip Peng Gao$^5$ \enskip Haoyu Wang$^3$}
\author{Xusheng Xiao$^{4,2}$ \enskip Bihan Wen$^6$ \enskip Yih-Chun Hu$^{1,2}$}

\affiliation{%
{	
	$^1$University of Illinois at Urbana-Champaign \quad
	$^2$\sys Consortium
}
}
\affiliation{ 
{	
		$^3$Beijing University of Posts and Telecommunications \quad
		$^4$Case Western Reserve University
}
}
\affiliation{ 
	{	
		$^5$University of California, Berkeley \quad
		$^6$Nanyang Technological University 
	}
}

\email{hyperservice.team@gmail.com}

\begin{abstract}\label{sec:abs}
	Blockchain interoperability, which allows state transitions 
across different blockchain networks, is critical functionality to facilitate major blockchain adoption. 
Existing interoperability protocols mostly focus on  
atomic token exchange between blockchains. However, as blockchains 
have been upgraded from passive distributed ledgers into programmable 
state machines (thanks to smart contracts), the scope of 
blockchain interoperability goes beyond just token exchange. 
In this paper, we present \sys, the first platform that delivers 
\emph{interoperability} and \emph{programmability} across \emph{heterogeneous} 
blockchains. \sys is powered by two innovative designs: 
\first a developer-facing programming framework that allows 
developers to build cross-chain applications in a unified programming 
model; and \second a secure blockchain-facing cryptography protocol that 
provably realizes those applications on blockchains. 
We implement a prototype of \sys in approximately 35,000 lines of code 
to demonstrate its practicality. Our experiment results show that \first 
\sys imposes reasonable latency, in order of seconds, on the end-to-end 
execution of cross-chain applications; \second the \sys platform 
is scalable to continuously incorporate additional production blockchains. 
\end{abstract}



\copyrightyear{2019} 
\acmYear{2019} 
\acmConference[CCS '19]{2019 ACM SIGSAC Conference on Computer and Communications Security}{November 11--15, 2019}{London, United Kingdom}
\acmBooktitle{2019 ACM SIGSAC Conference on Computer and Communications Security (CCS '19), November 11--15, 2019, London, United Kingdom}
\acmPrice{15.00}
\acmDOI{10.1145/3319535.3355503}
\acmISBN{978-1-4503-6747-9/19/11}

\maketitle

\section{Introduction}\label{sec:intro}
Over the last few years, we have witnessed  
rapid growth of several flagship blockchain applications, 
such as the payment system Bitcoin~\cite{btc} and the 
smart contract platform Ethereum~\cite{eth-whitepaper}. 
Since then, considerable effort has been 
made to improve the performance and security of individual blockchains, 
such as more efficient consensus algorithms~\cite{pbft-improve,dpos,bc-ng,monoxide},
improving transaction rate by sharding~\cite{rapidchain, omniledger, sharding-A, chainspace} and 
payment channels~\cite{payment-channel-A,payment-channel-B,payment-channel-C}, 
enhancing the privacy for smart contracts~\cite{hawk,sm-privacy,ekiden}, 
and reducing their vulnerabilities via program 
analysis~\cite{sm-analysis,sm-analysis-2,sm-analysis-3}.




As a result, in today's blockchain ecosystem, 
we see many distinct blockchains, 
falling roughly into the categories of public, private, and consortium 
blockchains~\cite{cmc}. In a world deluged with isolated blockchains, 
interoperability is power. Blockchain interoperability enables secure 
state transitions across different blockchains,
which is invaluable for connecting the
decentralized Web 3.0~\cite{vitalik_inter}. 
Existing interoperability proposals~\cite{chain_inter_A, chain_inter_B, chain_inter_C, pos_side} 
mostly center around atomic token exchange between two blockchains, 
aiming to eliminate the requirement of centralized exchanges.  
However, since smart contracts executing on blockchains have  
transformed blockchains from append-only distributed ledgers into programmable state 
machines, we argue that \emph{token exchange is not the complete scope of blockchain 
interoperability}. Instead, blockchain interoperability is complete only with \emph{programmability},
allowing developers to write decentralized applications executable across those 
disconnected state machines. 

We recognize at least two categories of challenges for simultaneously delivering 
programmability and interoperability. First, the programming model of 
cross-chain decentralized applications (or \dApps) is unclear. 
In general, from developers' perspective, it is desirable that 
cross-chain \dApps could preserve the same state-machine-based 
programming abstraction as single-chain contracts~\cite{ethereum-yellowpaper}. 
This, however, raises a virtualization challenge to abstract away the heterogeneity of 
smart contracts and accounts on different blockchains so 
that the interactions and operations among those contracts and accounts   
can be \emph{uniformly} specified when writing cross-chain \dApps. 

\begin{figure*}[t]
	\centering
	\mbox{
		\includegraphics[width=\textwidth]{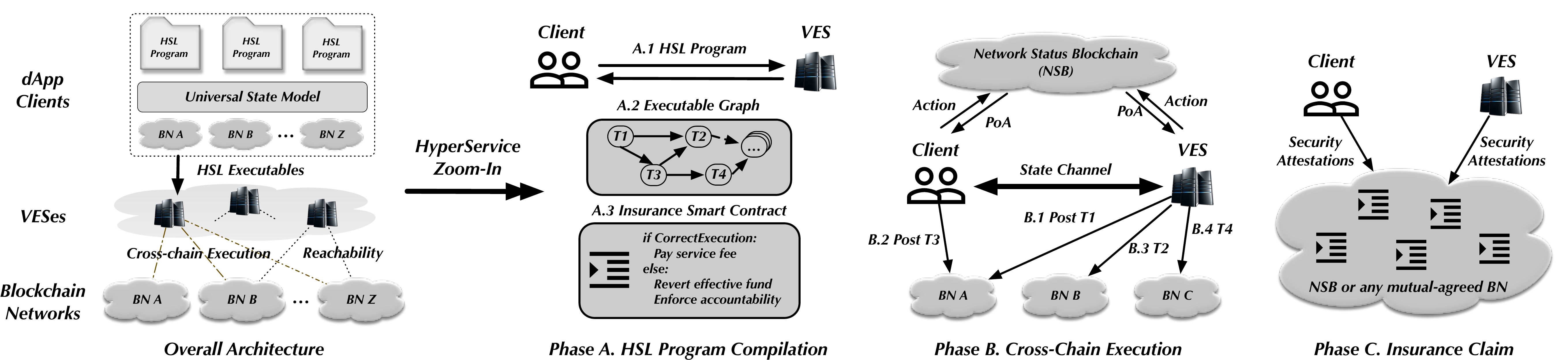}
	}
	\caption{The architecture of \sys.} \label{fig:arch}
\end{figure*}

Second, existing token-exchange oriented interoperability protocols, such as 
atomic cross-chain swaps (ACCS)~\cite{accs}, are not generic enough to 
realize cross-chain \dApps.  This is because the ``executables'' of those \dApps could 
contain more complex operations than token transfers. For instance, our example 
\dApp in \cref{sec:overview:hsl} invokes a smart contract using parameters 
obtained from smart contracts deployed on different blockchains. 
The complexity of this operation is far beyond mere token transfers. 
In addition, the executables of cross-chain \dApps often contain transactions 
on different blockchains, and the correctness of \dApps requires those 
transactions to be executed with certain preconditions and deadline constraints. 
Another technical challenge is to securely coordinate those transactions to enforce \dApp correctness
in a fully decentralized manner with zero trust assumptions.

To meet these challenges, we propose \sys, the first platform for building and executing \dApps 
across heterogeneous blockchains. At a very high level, \sys is powered by two innovative 
designs: a developer-facing \emph{programming framework} for writing cross-chain \dApps, 
and a blockchain-facing cryptography protocol to securely 
realize those \dApps on blockchains. 
Within this programming framework, we propose 
Unified State Model (\USM), a blockchain-neutral and extensible model 
to describe cross-chain \dApps, and the \HSL, 
a high-level programming language to write cross-chain 
\dApps under the \USM programming model. 
\dApps written in \HSL are further compiled into \sys executables which shall be  
executed by the underlying cryptography protocol. 

\UIP (short for universal inter-blockchain protocol) is the cryptography protocol 
that handles the complexity of cross-chain execution. 
\UIP is \first \emph{generic}, operating on any blockchain 
with a public transaction ledger, \second \emph{secure}, the executions of \dApps 
either finish with verifiable correctness or abort due
to security violations, where misbehaving parties are held accountable, 
and \third \emph{financially atomic}, meaning all involved parties experience almost zero financial losses, 
regardless of the execution status of \dApps. 
\UIP is fully trust-free, assuming no trusted entities. 

\parab{Contributions.} To the best of our knowledge, \sys is the first 
platform that simultaneously offers \emph{interoperability} and 
\emph{programmability} across \emph{heterogeneous} blockchains.
Specifically, we make the following major contributions in this paper. 

\first We propose the first programming framework for 
developing cross-chain \dApps. The framework greatly 
facilitates \dApp development by providing a virtualization 
layer on top of the underlying heterogeneous blockchains, yielding a unified 
model and a high-level language to describe and program \dApps. 
Using our framework, a developer can easily 
write cross-chain \dApps without implementing any cryptography. 

\second We propose \UIP, the first generic blockchain interoperability 
protocol whose design scope goes beyond cross-chain 
token exchanges. Rather, \UIP is capable of securely realizing complex cross-chain 
operations that involve smart contracts deployed on heterogeneous blockchains. 
We express the security properties of \UIP via an ideal functionality 
\fbip and rigorously prove that \UIP realizes 
\fbip in the Universal Composability (UC) framework~\cite{UC}. 

\third We implement a prototype of \sys in approximately 35,000 lines of code, 
and evaluate the prototype with three categories of cross-chain \dApps. 
Our experiments show that the end-to-end \dApp execution latency
imposed by \sys is in the order of seconds, and the \sys platform has sufficient capacity 
to continuously incorporate new production blockchains. 

\section{\sys Overview}\label{sec:overview} 

\subsection{Architecture}\label{sec:overview:arch}
As depicted in \cref{fig:arch}, 
architecturally, \sys 
consists of four components. 
\first \emph{\dApp Clients} are the gateways for \dApps to 
interact with the \sys platform. 
When designing \sys, we intentionally make clients to be lightweight, 
allowing both mobile and web applications to interact with \sys.
\second \emph{Verifiable Execution Systems (\VENs)}   
conceptually work as \emph{blockchain drivers} that 
compile the high-level \dApp programs given by the \dApp clients 
into blockchain-executable transactions, which are 
the runtime executables on \sys. \VENs and \dApp clients 
employ the underlying \UIP cryptography protocol to 
securely execute those transactions across different blockchains. 
\UIP itself has two building blocks: \third the Network Status Blockchain (\NSB) 
and \forth the Insurance Smart Contracts (\ISCs). The \NSB, conceptually, 
is a \emph{blockchain of blockchains}  
designed by \sys to provide an objective and unified view of the \dApps'
execution status, based on which the \ISCs 
arbitrate the correctness or violation of \dApp executions in a trust-free manner. 
In case of exceptions, the \ISCs financially revert 
all executed transactions to guarantee financial atomicity and 
hold misbehaved entities accountable. 

\subsection{Universal State Model}\label{sec:overview:usm}
\begin{table*}[t]
	\caption{Example of entities, operations and dependencies in \USM}\label{table:usm}
	\resizebox{\textwidth}{!}{%
	\begin{tabular}{|c|l||c|l||c|}
		\hline
		\textit{Entity Kind} & \multicolumn{1}{c||}{Attributes} & \textit{Operation Kind} & \multicolumn{1}{c||}{Attributes} & \textit{Dependency Kind} \\ \hline
		\haccount & address, balance, unit & \hpay & from, to, value, exchange rate & \hprecon \\ \hline
		\hcontract & address, state variables{[}{]}, interfaces{[}{]}, source & \hinvo & interface, parameters{[}const, Contract.SV, ...{]}, invoker & \hddl \\ \hline
	\end{tabular}
}
\end{table*}

A blockchain, together with smart contracts (or \dApps) executed on the blockchain, 
is often perceived as a state machine~\cite{ethereum-yellowpaper}. 
We desire to preserve the similar abstraction
for developers when writing cross-chain \dApps. Towards this end, we propose Unified State Model (\USM), 
a blockchain-neutral and extensible model for describing state transitions 
across different blockchains, which in essential defines cross-chain \dApps. 
\USM realizes a virtualization layer to unify the underlying heterogeneous blockchains. 
Such virtualization includes: \first blockchains, 
regardless of their implementations (\eg consensus mechanisms, 
smart contract execution environment, programming languages, and so on), 
are abstracted as \emph{objects} with public state variables and functions; 
\second developers program \dApps by specifying desired 
operations over those objects, along with the relative ordering 
among those operations, as if all the objects were local to a single machine. 
 
Formally, \USM is defined as $\mathcal{M} = \{\mathcal{E}, \mathcal{P}, \mathcal{C}\}$ where $\mathcal{E}$ is a set of \emph{entities}, $\mathcal{P}$ is a set of \emph{operations} performed over those entities, and $\mathcal{C}$ is a set of constraints defining the \emph{dependencies} of those operations. Entities are to describe the objects abstracted from blockchains. All entities are conceptually local to $\mathcal{M}$, regardless of which blockchains they are obtained from. Entities come with \emph{kinds}, and each entity kind has different attributes.
The current version of \USM defines two concrete kinds of entities, \emph{accounts} and \emph{contracts}, as tabulated in \cref{table:usm} (we discuss the extensions of \USM in \S~\ref{discussion:hsl}). 
Specifically, an account entity is associated with a uniquely identifiable address, as well as its balance in certain units. A contract entity, besides its address, is further associated with a list of public attributes, such as 
state variables, callable interfaces, and its source code deployed on blockchains. Entity attributes are crucial to 
enforce the security and correctness of \dApps during 
compilation, as discussed in \S~\ref{sec:overview:hsl}. 

An operation in \USM defines a step of 
computation performed over several entities. \cref{table:usm} lists two kinds of operations in \USM: a \emph{payment} operation that describes the balance updates between two account entities at a certain exchange rate; an \emph{invocation} operation that describes the execution of a method specified by the interface of a contract entity using compatible parameters, whose values may be obtained from other contract entities' state variables.  

Although operations are conceptually local, each operation is eventually compiled into one or more transactions on different blockchains, whose consensus processes are not synchronized. To honor the possible dependencies among events in distributed computing~\cite{clocks}, \USM, therefore, defines constraints to specify dependencies among operations. Currently, \USM supports two kinds of dependencies: \emph{preconditions} and \emph{deadlines}, where an operation can proceed only if all its preconditioning operations are finished, and an operation must be finished
within a bounded time interval after its dependencies are satisfied. 
Preconditions and deadlines offer desirable programming abstraction for \dApps: \first 
preconditions enable developers to organize their 
operations into a directed acyclic graph, where the state of upstream 
nodes is persistent and can be used by downstream nodes; 
\second deadlines are crucial to ensure the forward progress of \dApp executions. 

\subsection{\sys Programming Language}\label{sec:overview:hsl}
To demonstrate the usage of \USM, we develop \HSL, a programming 
language to write cross-chain \dApps under
\USM. 

\subsubsection{An Introductory Example for \HSL Programs} 
Financial derivatives are among the most commonly cited 
blockchain applications. Many financial derivatives rely on 
authentic data feed, \ie an \emph{oracle}, as inputs. 
For instance, a standard call-option contract needs a genuine strike price. 
Existing oracles~\cite{TC,rhombus} require a smart contract on the 
blockchain to serve as the front-end to interact with other client smart contracts. 
As a result, it is difficult to build a dependable and unbiased oracle that is simultaneously 
accessible to multiple blockchains, 
because we cannot simply deploy an oracle smart contract on each individual blockchain 
since synchronizing the execution of those oracle contracts requires blockchain interoperability, 
\ie we see a chicken-and-egg problem. 
This limitation, in turn, prevents 
\dApps from spreading their business across multiple blockchains. 
For instance, a call-option contract deployed on Ethereum forces 
investors to exercise the option using Ether, but not in other cryptocurrencies. 

As an introductory example, we shall see  
how conceptually simple, yet elegant, it is, from developers' perspective, 
to build a universal call-option \dApp that allows investors to 
natively exercise options with the cryptocurrencies they prefer. 
The code snippet shown in \cref{fig:hsl_program} is 
the \HSL implementation for the referred \dApp. 
In this \dApp, both Option contracts deployed 
on blockchains \hchainy and \hchainz rely on the same Broker contract on \hchainx 
to provide the genuine strike price (lines \ref{line:c2.settle} and 
\ref{line:c3.settle} in \cref{fig:hsl_program}). 
Detailed \HSL grammar is given in \cref{bnf:parser}. 
 
\begin{figure}
	\small
	\fbox {
		\parbox{0.95\columnwidth}{
			\begin{enumerate}[label=\textbf{\scriptsize \arabic*}, leftmargin=*,itemsep=0.25ex]
				\item {\color{comment} \# Import the source code of contracts written in different languages.}
				\item \himport(``broker.sol'', ``option.vy'', ``option.go'')
				\item {\color{comment} \# Entity definition. }
				\item {\color{comment} \# Attributes of a contract entity are implicit from its source code. }
				\item \haccount a1 = \hchainx::Account(0x7019..., 100, \hxcoin)
				\item \haccount a2 = \hchainy::Account(0x47a1..., 0, \hycoin)
				\item \haccount a3 = \hchainz::Account(0x61a2..., 50, \hzcoin)
				\item \hcontract c1 = \hchainx::Broker(0xbba7...)
				\item \hcontract c2 = \hchainy::Option(0x917f...)
				\item \hcontract c3 = \hchainz::Option(0xefed...)
				\item {\color{comment} \# Operation definition. }
				\item \hop op1 \hinvo c1.GetStrikePrice() \husing a1
				\item \hop op2 \hpay 50 \hxcoin \hfrom a1 \hto a2 \hwith 1 \hxcoin \has 0.5 \hycoin
				\item \hop op3 \hinvo c2.CashSettle(10, c1.StrikePrice) \husing a2 \label{line:c2.settle}
				\item \hop op4 \hinvo c3.CashSettle(5, c1.StrikePrice) \husing a3 \label{line:c3.settle}
				\item {\color{comment} \# Dependency definition. }
				\item op1 \hbefore op2, op4; op3 \hafter op2
				\item op1 \hddl 10 blocks; op2, op3 \hddl default; op4 \hddl 20 mins
			\end{enumerate}
		}
	}
	\caption{A cross-chain Option \dApp written in \HSL. 
	}\label{fig:hsl_program}
\end{figure}

\subsubsection{\HSL Program Compilation}\label{sec:overview:hsl_compiler}
The core of \sys programming framework is the \HSL compiler. 
The compiler performs two major tasks: \first enforcing 
security and correctness checks on \HSL programs and \second 
compiling \HSL programs into blockchain-executable transactions. 

One of the key differentiations of \sys is that it allows \dApps to natively define 
interactions and operations among  
smart contracts deployed on heterogeneous blockchains.
Since these smart contracts could be written in different languages,
\HSL provides a multi-language front end to 
analyze the source code of those smart contracts. It extracts 
the type information of their public state variables and functions, and then 
converts them into the unified types defined by \HSL (\cref{subsec:unifiedtypes}). 
This enables effective correctness checks on the \HSL programs (\cref{subsec:validation}). 
For instance, it ensures that all the parameters used in a contract \hinvo operation  
are compatible and verifiable, even if these arguments are extracted 
from remote contracts written in languages different from that of the invoking contract. 

Once a \HSL program passes the syntax and correctness checks, the compiler will generate 
an \emph{executable} for the program. 
The executable is structured in the form of a Transaction 
Dependency Graph, which contains \first the complete information for computing  
a set of blockchain-executable transactions, \second the metadata of each 
transaction needed for correct execution, and \third the preconditions and deadlines of 
those transactions that honor the dependency constraints specified in the \HSL program 
(\cref{subsec:compilation}). 

In \sys, the Verifiable Execution Systems (\VENs)
are the actual entities that own the \HSL compiler and therefore 
resume the aforementioned compiler responsibilities. Because of this, 
\VENs work as \emph{blockchain drivers} that bridge our high-level programming 
framework with the underlying blockchains. Each \VEN is a distributed system providing 
trust-free service to compile and execute \HSL programs given by \dApp clients. 
\VENs are trust-free because their actions taken during \dApp executions are verifiable. 
Each \VEN defines its own service model, 
including its reachability (\ie the set of blockchains that 
the \VEN supports), service fees  
charged for correct executions, and insurance 
plans (\ie the expected compensation to \dApps if the \VEN's 
execution is proven to be incorrect). \dApps have full 
autonomy to select \VENs that satisfy their requirements. 
In \cref{discussion:csp}, we lay out our visions for \VENs. 

Besides owning the \HSL compiler, \VENs also participate in
the actual executions of \HSL executables, as discussed below. 

\subsection{Universal Inter-Blockchain Protocol (\UIP)}\label{sec:overview:uip}
To correctly execute a \dApp, all the transactions in its executable  
must be posted on blockchains for execution, and meanwhile 
their preconditions and deadlines are honored. 
Although this executing procedure is conceptually simple (thanks to the \HSL abstraction), 
it is very challenging to enforce correct executions in a fully trust-free manner where \first 
no trusted authority is allowed to coordinate the executions on different blockchains and \second no   
mutual trust between \VENs and \dApp clients are established. 

To address this challenge, \sys designs \UIP, a cryptography protocol 
between \VENs and \dApp clients to securely execute \HSL executables on blockchains. 
\UIP can work on any blockchain with public ledgers, imposing no additional 
requirements such as their consensus protocols and contract execution environment. 
\UIP provides strong security guarantees for executing \dApps such that 
\dApps are correctly executed only if the correctness is 
publicly verifiable by all stakeholders; otherwise, \UIP   
holds the misbehaving parties accountable, and 
financially reverts all committed transactions 
to achieve financial atomicity. 

\UIP is powered by two innovative designs: 
the Network Status Blockchain (\NSB) and the 
Insurance Smart Contract (\ISC). 
The \NSB is a blockchain designed by \sys to provide 
objective and unified views on the status of \dApp executions.  
On the one hand, the \NSB consolidates
the finalized transactions of all underlying 
blockchains into Merkle trees, providing 
unified representations for transaction status in form of verifiable Merkle proofs. 
On the other hand, the \NSB supports  
Proofs of Actions (PoAs), allowing both \dApp clients and \VENs 
to construct proofs to certify their actions taken during cross-chain 
executions. The \ISC is a code-arbitrator. 
It takes transaction-status proofs constructed from the 
\NSB as input to determine the correctness or violation of \dApp executions, 
and meanwhile uses action proofs to determine the accountable 
parities in case of exceptions. 

In \cref{sec:UIP:theory}, we define the security properties of \UIP via an 
ideal functionality and then rigorously prove that \UIP realizes the ideal  
functionality in UC-framework~\cite{UC}.

\subsection{Assumptions and Threat Model}\label{sec:overview:threat}
We assume that the cryptographic primitives and the consensus protocol 
of all underlying blockchains are secure so that each of them 
can have the concept of \emph{transaction finality}.
On Nakamoto consensus based blockchains (typically permissionless), this is achieved by assuming 
that the probability of blockchain reorganizations drops exponentially as 
new blocks are appended (\emph{common-prefix property})~\cite{bitcoin_difficulty}. 
On Byzantine tolerance based blockchains (usually permissioned), 
finality is guaranteed by signatures from a quorum of permissioned 
voting nodes. For a blockchain, if the \NSB-proposed definition of 
transaction finality for the blockchain is accepted 
by users and \dApps on \sys, the operation (or trust) model 
(\eg permissionless or permissioned) and consensus efficiency (\ie the 
latency for a transaction to become final) of the blockchain
have provably no impact on the security guarantees of our \UIP protocol. 
We also assume that each underlying blockchain has a public ledger that allows external parties 
to examine and prove transaction finality and the \emph{public} state 
of smart contracts. 

The correctness of \UIP relies on the correctness of the \NSB. 
An example implementation of \NSB is a permissioned blockchain, 
where any information on \NSB becomes legitimate only if a quorum of consensus nodes that maintain the \NSB 
have approved the information. We thus assume that at least $\mk$ consensus nodes 
of the \NSB are honest, where $\mk$ is the quorum threshold (\eg the majority). 
In this design,
an \NSB node is not required to become either a full or light node for any 
of the underlying blockchains. 

We consider a Byzantine adversary that interferes with our \UIP protocol arbitrarily, 
including delaying and reordering network messages indefinitely, and  
compromising protocol participants.
As long as at least one protocol participant is not
compromised by the adversary,
the security properties of \UIP are guaranteed.

\section{Programming Framework}\label{sec:hsl}

\begin{figure}[t]
	\centering
	\mbox{
		\includegraphics[width=0.95\columnwidth]{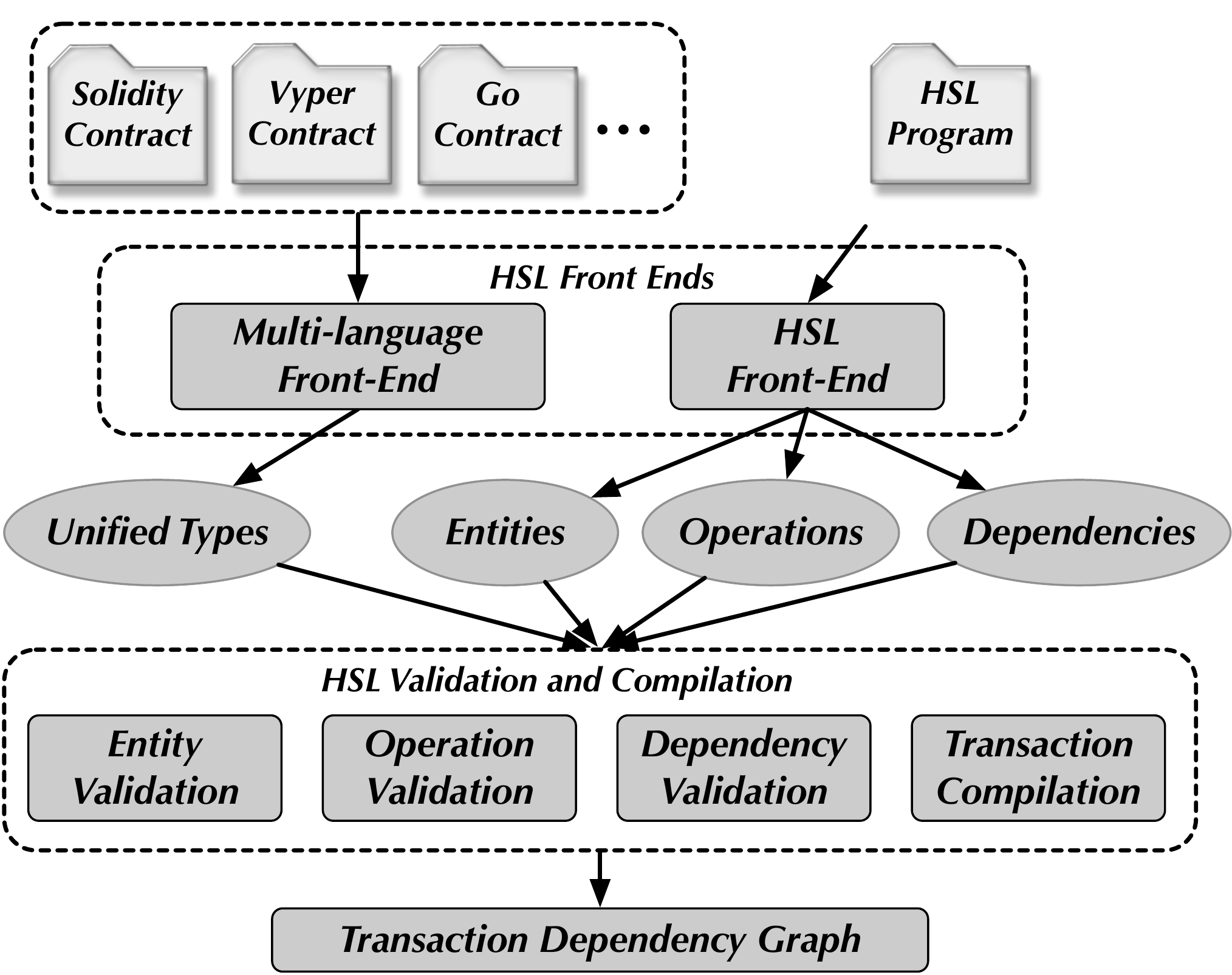}
	}
	\caption{Workflow of \HSL compilation.} \label{fig:hsl}
\end{figure}

The design of the \sys programming framework centers around the \HSL compiler.  
\cref{fig:hsl} depicts the compilation workflow. 
The \HSL compiler has two frond-ends: one for extracting 
entities, operations, and dependencies from a \HSL program and 
one for extracting public state variables and methods from 
smart contracts deployed on blockchains. 
A unified type system is designed to ensure that smart contracts written in 
different languages can be abstracted as interoperable entities defined in the 
\HSL program. Afterwards, the compiler performs semantic validations on all entities, 
operations and dependencies to ensure the security and correctness of the \HSL program.
Finally, the compiler produces an executable for the \HSL program, which is 
structured in the form of a transaction dependency graph.
We next describe the details of each component. 

\subsection{Unified Type System}\label{subsec:unifiedtypes}
The \USM is designed to provide a unified virtualization 
layer for developers to define \hinvo operations in their \HSL programs, without 
handling the heterogeneity of contract entities. 
Towards this end, the programming framework internally defines a 
Unified Type System so that state variables and methods of all 
contract entities can be abstracted using the unified types when 
writing \HSL programs. This enables the \HSL compiler to ensure that all arguments 
specified in an \hinvo operation are \emph{compatible} (\cref{subsec:validation}).  

Specifically, the unified type system defines nine elementary types, as shown in \cref{tab:uts}. 
Data types that are commonly used in smart contract programming languages 
will be \emph{mapped} to these unified types during compilation. 
For example, Solidity does not fully support fixed-point number, 
but Vyper ($decimal$) and Go ($float$) do. 
Also, Vyper's string is fixed-sized (declared via string$\left[Integer \right]$), 
but Solidity's string is dynamically-sized (declared as string).
Our multi-lang front-end recognizes these differences and performs type conversion to 
map all the numeric literals including integers 
and decimals to the $Numeric$ type, and the strings to the $String$ type.
For types that are similar in Solidity, Vyper, and Go, 
such as $Boolean$, $Map$, and $Struct$, we simply map them to the 
corresponding types in our unified type system. 
Finally, Solidity and Vyper provide special types for representing 
contract addresses, which are mapped to the $Address$ type.
But Go does not provide a type for contract addresses, and thus 
Go's $String$ type is mapped to the $Address$ type.
The mapping of language-specific types to the unified type system is 
tabulated in~\cref{tab:uts}. Our unified type system is horizontally scalable to support 
additional strong-typed programming languages. 
Note that the use of complex data types as contract function parameters has not been 
fully supported yet in production. We thus leave complex types 
in \HSL to future work. 

\begin{table}[t]
	\caption{Unified type mapping for Solidity, Vyper, and Go}
	\scriptsize
	\label{tab:uts}

	\begin{tabular}{|p{1cm}||p{1.9cm}||p{1.9cm}||p{1.9cm}|}
		\hline
		\textbf{\textit{Type} }& \textbf{\textit{Solidity}} & \textbf{\textit{Vyper}} 
		& \textbf{\textit{Go} } \\ \hline
		\textit{Boolean} & bool & bool & bool \\ \hline
		\textit{Numeric} & int, uint & int128, uint256, decimal, unit type & int, uint, uintptr, float \\ \hline
		\textit{Address} & address & address & string \\ \hline
		\textit{String} & string  & string & string \\ \hline
		\textit{Array} & array, bytes & array, bytes & array, slice \\ \hline
		\textit{Map} & mapping & map & map \\ \hline
		\textit{Struct} & struct & struct & struct \\ \hline
		\textit{Function} & function, enum & def & func \\ \hline
		\textit{Contract} & Contract & file & type \\ \hline
	\end{tabular}

\end{table}

\vspace{-0.2cm}
\subsection{\HSL Language Design}
\label{subsec:hsldesign}
The language constructs provided by \HSL are coherent with \USM, allowing  
developers to straightforwardly specify entities, operations, and dependencies in \HSL programs. 
One additional construct, \himport, is added to import    
the source code of contract entities, as discussed below. 
\cref{bnf:parser} shows the representative rules of \HSL. 
We omit the terminal symbols such as \emph{$\langle$id$\rangle$ }and \emph{$\langle$address$\rangle$}.

\parab{Contract Importing}.
Developers use the \emph{$\langle$import$\rangle$} rule to include 
the source code of contract entities. 
Depending on the programming language of an imported contract, 
\HSL's multi-lang front end uses the corresponding parser to parse the source code, 
based on which it performs semantic validation (\cref{subsec:validation}). 
For security purpose, the compiler should verify that the imported source code is consistent 
with the actual deployed code on blockchain, for instance, by comparing their compiled byte code. 

\parab{Entity Definition}.
The \emph{$\langle$entity\_def$\rangle$} rule specifies 
the definition of an \haccount or a \hcontract entity. 
An entity is defined via constructor, where 
the on-chain (\emph{$\langle$address$\rangle$}) of the entity is a required parameter. 
An \haccount entity can be initialized with an optional unit (\emph{$\langle$unit$\rangle$}) to specify 
the cryptocurrency held by the account. 
All \hcontract entities must have the corresponding contract objects/classes 
in one of the imported source code files.  
Each entity is assigned with a name (\emph{$\langle$entity\_name$\rangle$}) that 
can be used for defining operations.

\begin{BNF}[t]
	\footnotesize
	
	\begin{mdframed}
		
		\setlength{\grammarparsep}{-2pt} 
		\setlength{\grammarindent}{8em} 

		\begin{grammar}
			<hsl> ::= (<import>)+ (<entity\_def>)+ (<op\_def>)+ (<dep\_def>)*
		\end{grammar}
		
		\textbf{Contract Imports:}
		\begin{grammar}
			<import> ::= `import' `('  <file> (`,' <file>)* `)'
			
			<file> ::= <string>
			
		\end{grammar}
		
		\vspace*{1ex}

		\textbf{Entity Definition:}
		\begin{grammar}
			<entity\_def> ::= <entity\_type> <entity\_name> `=' <chain\_name> `::' <constructor>
			
			<entity\_name> ::= <id>
			
			<chain\_name> ::= `Chain' <id>
			
			<constructor> ::= <contract\_type> `('  <address>, (<unit>)? `)'
			
			<contract\_type> ::= `Account' | <id>
			
			<entity\_type> ::= `account' | `contract'
						
		\end{grammar}
		
		\vspace*{1ex}
		
		\textbf{Operation Definition:}
		\begin{grammar}
			<op\_def> ::= <op\_payment> | <op\_invocation>
			
			<op\_payment> ::= `op' <op\_name> `payment' <coin> <accts> <exchange>
			
			<op\_name> ::= <id>
			
			<coin> ::= <num> <unit>
			
			<accts> ::= `from' <acct> `to' <acct> 
			
			<acct> ::= <id>
			
			<exchange> ::= `with' <coin> `as'  <coin>
			
			<op\_invocation> ::= `op' <op\_name> `invocation' <call> `using' <acct>
			
			<call>  ::= <recv> `.' <method> `(' (arg)*`)'
			
			<arg> ::= <int> | <float> | <string> | <state\_var>
			
			<state\_var> ::= <varname> `.' <prop>
		\end{grammar}
		
		\vspace*{1ex}
		
		\textbf{Dependency Definition:}
		\begin{grammar}
			<dep\_def> ::=  <temp\_deps> | <del\_deps>
			
			<temp\_deps> ::= <temp\_dep> (`;' <temp\_dep>)*
			
			<temp\_dep> ::= <op\_name> (`before' | `after') <op\_name> (`,' <op\_name>)*
			
			<del\_deps> ::= <del\_dep> (`;' <del\_dep>)*
			
			<del\_dep> ::= <op\_name> (`,' <op\_name>)* `deadline' <del\_spec>
			
			<del\_spec> ::= <int> `blocks'| `default' | <int> <time\_unit>

		\end{grammar}

	\end{mdframed}
	
	\caption{Representative BNF grammar of \HSL}\label{bnf:parser}
\end{BNF}

\parab{Operation Definition}.
The \emph{$\langle$op\_def$\rangle$} rule specifies the definition of 
a \hpay or an \hinvo operation.
A \hpay operation (\emph{$\langle$op\_payment$\rangle$}) specifies the 
transfer of a certain amount of coins (\emph{$\langle$coin$\rangle$}) between two accounts that 
may live on different blockchains (\emph{$\langle$accts$\rangle$}).
Note that no new coins on any blockchains are ever created during the operation. 
The \emph{$\langle$exchange$\rangle$} rule is used to specify the 
exchange rate between the coins held by the two accounts.
An \hinvo operation (\emph{$\langle$op\_invocation$\rangle$}) 
specifies calling one contract entity's public method with certain arguments (\emph{$\langle$call$\rangle$}).
The arguments passed to a method invocation can be literals (\emph{$\langle$int$\rangle$}, \emph{$\langle$float$\rangle$}, \emph{$\langle$string$\rangle$}), and state variables (\emph{$\langle$state\_var$\rangle$}) of other contract entities. 
When using state variables, semantic validation is required (\cref{subsec:validation}). 

\parab{Dependency Definition}.
The \emph{$\langle$dep\_def$\rangle$} specifies the rule of defining 
preconditions and deadlines for operations. 
A \hprecon (\emph{$\langle$temp\_deps$\rangle$}) 
specifies the temporal constraints for the execution order of operations.
A \hddl (\emph{$\langle$del\_deps$\rangle$}) 
specifies the deadline constraints of each operation. The 
deadline dependency may be given either using the number of blocks 
on \NSB (\emph{$\langle$int$\rangle$ $blocks$}) or in absolute time 
(\emph{$\langle$int$\rangle$ $\langle$time\_unit$\rangle$}), as explained in \cref{subsec:compilation}. 

\subsection{Semantic Validation}\label{subsec:validation}
The compiler performs two types of semantic validation to 
ensure the security and correctness of \HSL programs. 
First, the compiler guarantees the \emph{compatibility} and \emph{verifiability} 
of the arguments used in \hinvo operations, especially when those    
arguments are obtained from other contract entities. 
For compatibility check, the compiler performs type checking 
to ensure the types of arguments and the types of method parameters are mapped to the same unified type.
For verifiability check, the compiler ensures that only literals and state 
variables that are publicly stored on blockchains are eligible to 
be used as arguments in \hinvo operations. 
For example, the return values of method calls to a contract 
entity are not eligible if these results are not persistent on blockchains. 
This requirement is necessary for the \UIP protocol  
to construct publicly verifiable attestations to prove that 
correct values are used to invoking contracts during actual on-chain execution. 
Second, the compiler performs dependency validation to make sure that the  
dependency constraints defined in a \HSL program uniquely specify a  
directed acyclic graph connecting all operations.
This ensures that no conflicting temporal constraints are specified. 

\subsection{\HSL Program Executables}\label{subsec:compilation}
\begin{figure}[t]
	\centering
	\mbox{
		\includegraphics[width=0.95\columnwidth]{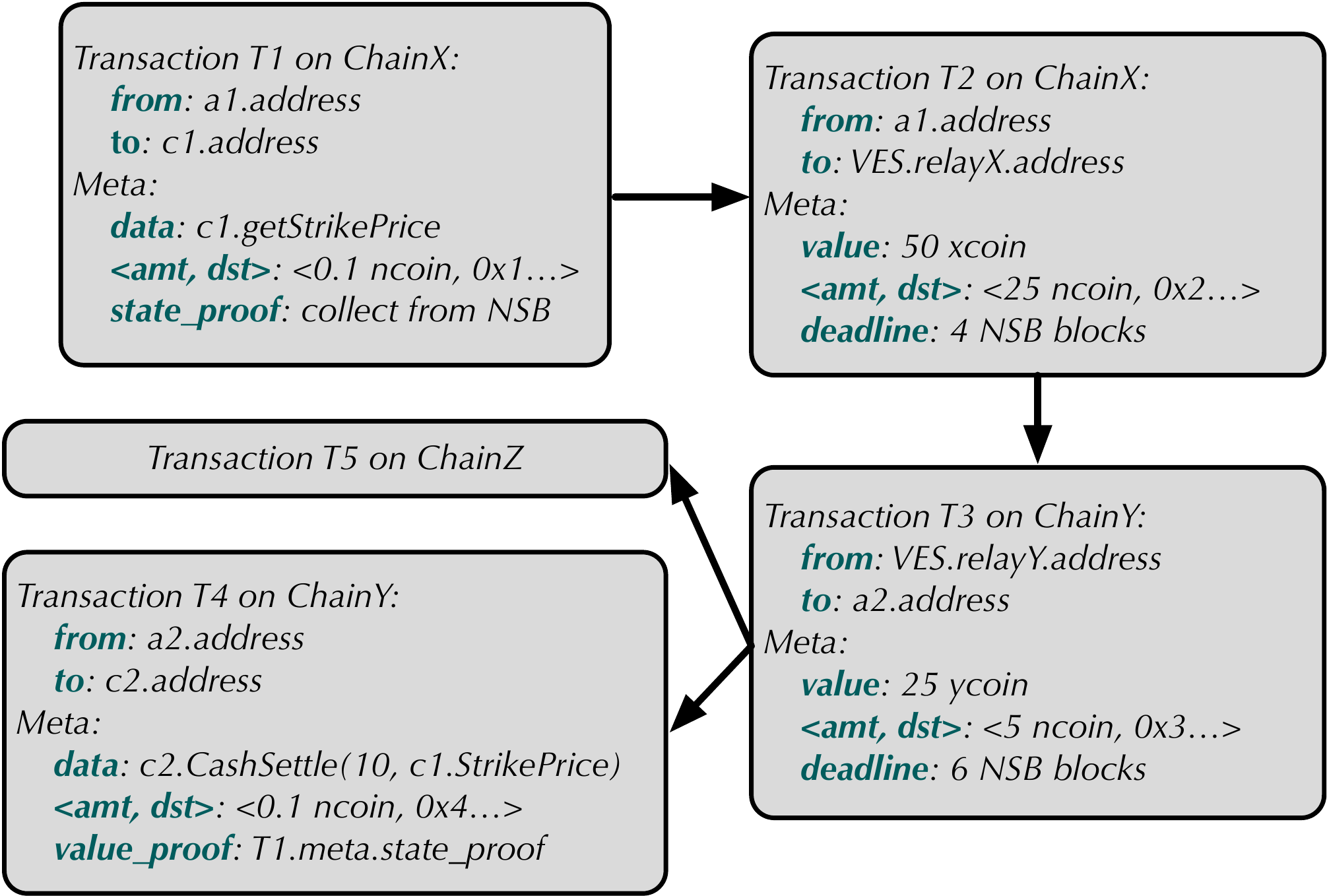}
	}
	\caption{\gt generated for the example \HSL program.} \label{fig:tdg}
\end{figure}

Once a \HSL program passes all validations, the \HSL compiler generates executables for 
the program in form of a transaction dependency graph \gt.
Each vertex of \gt, referred to as a \emph{transaction wrapper}, contains 
the complete information to compute an on-chain transaction 
executable on a specific blockchain, as well as additional metadata for the transaction. 
The edges in \gt define the preconditioning requirements among transactions, which 
are consistent with the dependency constraints specified by the \HSL program. 
\cref{fig:tdg} show the \gt generated for the \HSL program in 
\cref{fig:hsl_program}. 

A transaction wrapper is in form of $\mt := [\textsf{from}, \textsf{to}, \textsf{seq}, \hamt]$, 
where the pair <$\textsf{from}, \textsf{to}$> decides the 
sending and receiving addresses of the on-chain transaction, 
\textsf{seq} (omitted in \cref{fig:tdg}) represents the sequence number of $\mt$ in \gt, 
and \hamt stores the structured and customizable metadata for $\mt$. 
Below we explain the fields of \hamt. 
First, to achieve financial atomicity, \hamt must populate a tuple $\langle \amt, \dst \rangle$ for fund reversion. 
In particular, \amt specifies the total value that the $\emph{\textsf{from}}$ address 
has to spend when $\mt$ is committed on its destination blockchain, which includes 
both the explicitly paid value in $\mt$, as well as any gas fee. If the entire execution fails with exceptions 
whereas $\mt$ is committed, the \dst account is guaranteed to receive 
the amount of fund specified in \amt. As we shall see in \cref{sec:UIP:fisc}, 
the fund reversion is handled by the Insurance Smart Contract (\ISC). Therefore, 
the unit of \amt (represented as \hncoin in \cref{fig:tdg}) is given based on the cryptocurrency used by 
the blockchain where the \ISC is deployed, and the \dst should live on the hosting blockchain as well. 

Second, for a transaction (such as \emph{T1}) whose resulting state is subsequently 
used by other downstream transactions (such as \emph{T4}), 
its \hamt needs to be populated with a corresponding state proof. 
This proof should be collected from the transaction's destination blockchain after the transaction is finalized 
(\cf \cref{sec:fven:watching}). Third, a cross-chain payment operation in the \HSL program results in 
multiple transactions in \gt. For instance, to realize the $op1$ in \cref{fig:hsl_program}, 
two individual transactions, involving the \emph{relay accounts} owned by the \VEN, 
are generated. As blockchain drivers, each \VEN is supposed to own some accounts on 
all blockchains that it has visibility so that the \VEN is able to send and receive transactions 
on those blockchains. For instance, in \cref{fig:tdg}, the \emph{\textsf{relayX}} 
and \emph{\textsf{relayY}} are two accounts used by the \VEN to bridge  
the balance updates between  \emph{\textsf{ChainX::a1}} and \emph{\textsf{ChainY::a2}}. 
Because of those \VEN-owned accounts, \gt is in general \VEN-specific. 

Finally, the deadlines of transactions could be specified    
using the number of blocks on the \NSB. This is because the \NSB constructs 
a unified view of the status of all underlying blockchains and therefore can measure 
the execution time of each transaction. Specifically, the deadline of 
a transaction $\mt$ is measured as the number of blocks between two \NSB blocks 
$\mb_1$ and $\mb_2$ (including $\mb_2$), where $\mb_1$ proves the finalization 
of $\mt$'s last preconditioned transaction and $\mb_2$ proves the finalization 
of $\mt$ itself. We explain in detail how the finality proof is constructed based 
on \NSB blocks in \cref{sec:fven:exchange}. Transaction deadlines are indeed 
enforced by the \ISC using the number of \NSB blocks. 
Note that to improve expressiveness, the \HSL language also allows 
developers to define deadlines in time intervals (\eg minutes). The compiler 
will then convert those time intervals into numbers of \NSB blocks. 

In summary, the executable produced by the \HSL complier defines the blueprint 
of cross-blockchain execution to realize the \HSL program. 
It is the input instructions that direct the underlying cryptography protocol \UIP, as detailed below. 
\section{\UIP Design Detail}\label{sec:UIP}

\UIP is the cryptography protocol that executes \HSL program executables. 
The main protocol \pbip is divided into \emph{five} preliminary protocols. 
In particular, \fven and \fdapp define the execution protocols implemented 
by \VENs and \dApp clients, respectively. 
\fnsb and \fisc are the protocol realization of the \NSB and \ISC, respectively. 
Lastly, \pbip includes \pbc, the protocol realization of a 
general-purposed blockchain. 
Overall, \pbip has two phases: the execution phase where 
the transactions specified in the \HSL executables are posted on blockchains 
and the insurance claim phase where the execution correctness or violation 
is arbitrated.  

\subsection{Protocol Preliminaries}
\subsubsection{Runtime Transaction State}\label{sec:uip:state}
During the execution phase, a transaction may be 
in any of the following state $\{\unknown, \init, \initd, \open, \opend, \close\}$, 
where a latter state is considered more \emph{advanced} than a former one. 
The state of each transaction must be gradually promoted following the above sequence.  
For each state (except for the \unknown), \pbip produces a corresponding attestation 
to prove the state. When the execution phase terminates, the final execution status of the \HSL program is 
collectively decided by the state of all transactions, based on which \fisc arbitrates 
its correctness or violation. 


\subsubsection{Off-Chain State Channels}\label{sec:fven:channel}
The protocol exchange between \fven and \fdapp can be conducted via 
off-chain state channels for low latency. One challenge, however, is 
that it is difficult to enforce accountability for non-\close transactions without 
preserving the execution steps by both parties.  
To address this issue, \pbip proposes Proof of Actions (PoAs), allowing \fven and \fdapp to 
stake their execution steps on \NSB. As a result, the \NSB is treated as a 
publicly-observable \emph{fallback} communication medium for the off-chain channel. 
The benefit of this dual-medium design is that the protocol exchange 
between \fven and \fdapp can still proceed agilely via off-chain channels in typical scenarios, 
whereas the full granularity of their protocol exchange is preserved on the \NSB in case of exceptions, 
eliminating the ambiguity for accountability enforcement. 

As mentioned in \cref{sec:uip:state}, \pbip produces 
security attestations to prove the runtime state of transactions. 
As we shall see below, an attestation may come in two forms: a certificate, denoted by \atte, signed 
by \fven or/and \fdapp during their off-chain exchange, or an \emph{on-chain} Merkle proof, denoted 
by \merk, constructed using the \NSB and underlying blockchains. An \atte and its corresponding 
\merk are treated equivalently by the \fisc in code arbitration. 

\subsubsection{Architecture of the \NSB}\label{sec:uip:nsb}
\begin{figure}[t]
	\centering
	\mbox{
		\includegraphics[width=0.99\columnwidth]{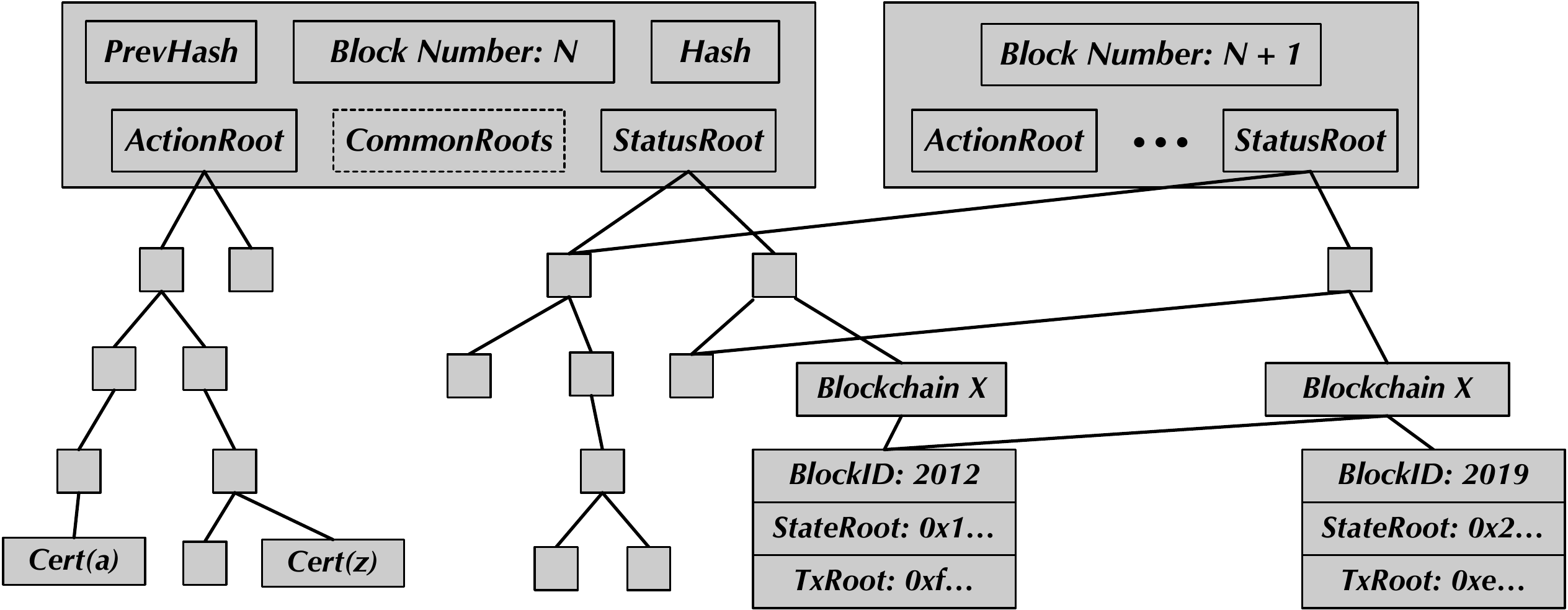}
	}
	\caption{The architecture of \NSB blocks.} \label{fig:nsb_block}
\end{figure}

The \NSB is a blockchain designed to provide an objective 
view on the execution status of \dApps. \cref{fig:nsb_block} depicts the 
architecture of \NSB blocks. Similar to typical blockchain blocks, an \NSB block 
contains several common fields, such as the hash fields to link blocks together and the   
Merkle trees to store transactions and state. To support the extra 
functionality of the \NSB, an \NSB block contains two additional Merkle tree roots: 
\textsf{StatusRoot} and \textsf{ActionRoot}. 

\textsf{StatusRoot} is the root of a Merkle tree (referred as \stmt) that
stores \emph{transaction status} of underlying blockchains. 
The \NSB represents the transaction status of a blockchain based on the 
\textsf{TxRoots} and \textsf{StateRoots} retrieved from the blockchain's public ledger. 
Although the exact namings may vary on different blockchains, in general, the \textsf{TxRoot} and 
\textsf{StateRoot} in a blockchain block represent the root of a Merkle tree storing transactions and 
storage state (\eg account balance, contract state), respectively. 
Note that the \NSB only stores \emph{relevant} blockchain state, where 
a blockchain block is considered to be relevant if the block packages at 
least one transaction that is part of any \dApp executables.

\textsf{ActionRoot} is the root of a Merkle tree (referred to as \acmt)
whose leaf nodes store certificates computed by \VENs and \dApp clients. 
Each certificate represents a certain step taken by either the \VEN or the \dApp 
client during the execution phase. To prove such an action, a party needs  
to construct a Merkle proof to demonstrate that the certificate mapped to the action 
can be linked to a committed block on the \NSB. These PoAs are crucial for the \ISC to 
enforce accountability if the execution fails. 
Since the information of each \acmt is static, we lexicographically 
sort the \acmt to achieve fast search and convenient proof of 
non-membership.

Note that the construction of \stmt ensures that each underlying blockchain can 
have a dedicated subtree for storing its transaction status. This makes the \NSB 
\emph{shardable on the granularity of individual blockchains}, ensuing that the \NSB is horizontally 
scalable as \sys continuously incorporates additional  
blockchains. \fnsb, discussed in \cref{sec:UIP:fnsb}, is the protocol that specifies the detailed
construction of both roots and guarantees their correctness. 

\newcommand{\add}{\textsf{Add}}
\newcommand{\Daemon}{\textbf{\textsf{Daemon}}\xspace}
\newcommand{\Init}{\textbf{\textsf{Init:}}\xspace}
\newcommand{\storage}{\textsf{Data}}	
\newcommand{\now}{\textsf{now}}
\newcommand{\from}{\textsf{from}}
\newcommand{\trans}{\textsf{trans}}
\newcommand{\resp}{\emph{\textsf{resp}}\xspace}
\newcommand{\Ag}{\textsf{A}_\revs\xspace}
\renewcommand{\st}{\textsf{T}_{\textsf{state}}\xspace}
\newcommand{\tid}{\textsf{tid}\xspace}
\newcommand{\map}{\textsf{map}\xspace}
\newcommand{\status}{\textsf{state}\xspace}
\newcommand{\dest}{\textsf{dest}\xspace}

\begin{figure*}
  \small
  \resizebox{\textwidth}{!}{
	\fbox {
		\parbox{1.005\textwidth}{
			\vspace*{-0.15in}
			\begin{multicols}{2}
			\begin{enumerate}[label=\textbf{\scriptsize \arabic*}, leftmargin=*,itemsep=0.25ex]
				\item \Init $\storage := \emptyset$
				\item \Daemon \pcmpl(): 
				\item \quad generate the session ID \sid $\leftarrow \{0, 1\}^\lambda$ 
				\item \quad call $[\cid,  \contract]:=~$\fisc.$\xcrcb(\rgt)$ 
				\item \quad send $\cert([\sid, \rgt, \contract]; \sigv)$ to \fdapp for approval 
				\item \quad halt until $\cert([\sid, \rgt, \contract]; \sigv, \sigd)$ is received
				\item \quad package \contract as a valid transaction $\widehat{\contract}$ 
				\item \quad call \fnsb.$\exec(\widehat{\contract})$ to deploy the $\widehat{\contract}$
				\item \quad halt until $\widehat{\contract}$ is initialized on \fnsb
				\item \quad call \fisc.$\xskfb$ to stake the required funds in \fisc 
				\item \quad halt until $\md$ has staked its required funds in \fisc 
				\item \quad initialize $\storage[\sid] := \{ \rgt, \cid, S_{\atte}{=}\emptyset, S_{\merk}{=}\emptyset \}$
				
				\item \Daemon \probe(\sid, \{\pbc, ...\}) \private: 
				\item \quad $(\rgt, \_, S_{\atte}, S_{\merk}) := \storage[\sid]$; 
					abort if not found 
				\item \quad \textbf{for each} $\mt \in \rgt$ \bco 
				\item \quad \quad \textbf{continue} if $\mt.\status$ is not \opend 
				\item \quad \quad identify $\mt$'s on-chain counterpart $\wtt$ 
				\item \quad \quad 
					\textbf{continue} if \pbc.$\textsf{Status}(\wtt)$ is not \textsf{committed}
				\item \quad \quad get $\tsclose := $ \fnsb.$\xbhb()$
				\item \quad \quad  compute 
					$C_\close^\mt := \cert([\wtt, \close, \sid, \mt, \tsclose], \sigv)$ 
				\item \quad \quad call \fdapp.$\xcltb(C_\close^\mt)$ to negotiate the \close attestation 
				\item \quad \quad call \pbc.$\xmkpb(\wtt)$ to obtain a finalization proof for $\wtt$
				\item \quad \quad denote the finalization proof as $\merk_\mt^{c_1}$ (\cref{fig:merkle_proof}) 
				\item \quad \quad update $S_{\atte}.\add(C_\close^{\mt})$ and  $S_{\merk}.\add(\merk_\mt^{c_1})$
				
				\item \Daemon \probe(\sid, \fnsb) \private: 
				\item \quad $(\rgt, \_, S_{\atte}, S_{\merk}) := \storage[\sid]$; abort if not found 
				\item \quad watch four types of attestations 
					\{$\atte^{\id}, \atte^{o}, \atte^{\od}, \atte^{c}$\}
				\item \quad process \emph{fresh} attestations via corresponding handlers (see below) 
				\item \quad {\color{comment} \# Retrieve alternative attestations if necessary.}
				\item \quad \textbf{for each} $\mt \in \rgt$ \bco 
				\item \quad \quad \textbf{if} $\mt.\status = \opend$ \textbf{and} $\merk_\mt^{c_1} \in S_{\merk}$ \bco 
				\item \quad \quad \quad retrieve the roots $[R,...]$ of the proof $\merk_\mt^{c_1}$ 
				\item \quad \quad \quad call \fnsb.$\xmkpb([R,...])$ to obtain a status proof $\merk_\mt^{c_2}$
				\item \quad \quad \quad \textbf{continue} if $\merk_\mt^{c_2}$ is not available yet on \fnsb
				\item \quad \quad \quad compute the complete
				proof $\merk_\mt^c := [\merk_\mt^{c_1}, \merk_\mt^{c_2}]$ 
				\item \quad \quad \quad update $\mt.\status := \close$ and $S_{\merk}.\add(\merk_\mt^c)$ 
				\item \quad compute eligible transaction set $\ms$ using the current state of \gt 
				\item \quad \textbf{for each} $\mt \in \ms$: 
				\item \quad \quad \textbf{continue} if  $\mt.\status$ is not \unknown 
				\item \quad \quad \textbf{if} $\mt.\textsf{from} = $ \fdapp: 
				\item \quad \quad \quad compute $\atte_\mt^i := \cert([\mt, \init, \sid]; \sigv)$ 
				\item \quad \quad \quad call \fdapp.$\xinittb(\atte_\mt^i)$ to request initialization
				\item \quad \quad \quad call \fnsb.$\xaddab(\atte_\mt^i)$ to prove $\atte_\mt^i$ is sent 
				\item \quad \quad \quad update $S_{\atte}.\add(\atte_\mt^i)$ and $\mt.\status := \init$
				\item 
					\noindent \adjustbox{bgcolor=gray!20,minipage=[t]{\linewidth}}{
					\quad \quad \quad non-blocking wait until \fnsb.$\xmkpb(\atte_\mt^{i})$ 
					rt. $\merk_\mt^{i}$
					\item \quad \quad \quad update $S_{\merk}.\add(\merk_\mt^{i})$
				}
				\item \quad \quad \textbf{else}: call $\emph{self}.\xsditedb(\sid, \mt)$
				
				\item \textbf{Upon Receive} \sdited(\sid, $\mathcal{T}$) \private: 
					\hspace*{\fill} \textcolor{olive}{\emph{Northbound}}
				\item \quad $(\rgt, \_, S_{\atte}, S_{\merk}) := \storage[\sid]$; abort if not found 
				\item \quad compute and sign the on-chain counterpart $\wtt$ for $\mt$ 
				\item \quad compute $\atte_\mt^{\id} := \cert([\wtt, \initd, \sid, \mt]; \sigv)$ 
				\item \quad call \fdapp.$\xitdtb(\atte_\mt^{\id})$ to request opening of initialized $\mt$  
				\item \quad call \fnsb.$\xaddab(\atte_\mt^{\id})$ to prove $\atte_\mt^{\id}$ is sent 
				\item \quad update $S_{\atte}.\add(\atte_\mt^{\id})$ and $\mt.\status := \initd$ 
				\item
				\noindent \adjustbox{bgcolor=gray!20,minipage=[t]{\linewidth}}{
					\quad non-blocking wait until \fnsb.$\xmkpb(\atte_\mt^{\id})$ 
					returns $\merk_\mt^{\id}$
					\item \quad update $S_{\merk}.\add(\merk_\mt^{\id})$
				}
				
				\item \textbf{Upon Receive} \rvited($\atte_\mt^\id$) \public:
					\hspace*{\fill}\textcolor{olive}{\emph{Southbound}}
				\item \quad assert $\atte_\mt^\id$ has the valid 
					form of $\cert([\wtt, \initd, \sid, \mt]; \sigd)$ 
				\item \quad $(\_, \_, S_{\atte}, S_{\merk}) := \storage[\sid]$; abort if not found 
				\item \quad abort if the $\atte_\mt^i$ corresponding 
					to $\atte_\mt^\id$ is not in $S_{\atte}$  
				\item \quad assert $\wtt$ is correctly associated with the wrapper $\mt$ \label{line:associate}
				\item \quad get $\tsopen := $ \fnsb.$\xbhb()$ 
				\item \quad compute $\atte_\mt^{o} := \cert([\wtt, \open, \sid, \mt, \tsopen]; \sigv)$
				\item \quad call \fdapp.$\xoptb(\atte_\mt^{o})$ to request opening for $\mt$
				\item \quad call \fnsb.$\xaddab(\atte_\mt^{o})$ to prove $\atte_\mt^{o}$ is sent 
				\item \quad update $S_{\atte}.\add(\atte_\mt^{o})$ and $\mt.\status := \open$
				\item
				\noindent \adjustbox{bgcolor=gray!20,minipage=[t]{\linewidth}}{
					\quad non-blocking wait until \fnsb.$\xmkpb(\atte_\mt^{o})$ 
						returns $\merk_\mt^{o}$ 
					\item \quad update $S_{\merk}.\add(\merk_\mt^{o})$ 
				}
				
				\item \textbf{Upon Receive} \opt($\atte_T^{o}$) \public: 
					\hspace*{\fill}\textcolor{olive}{\emph{Northbound}}
				\item \quad assert $\atte_T^{o}$ has valid form of 
					$\cert([\wtt, \open, \sid, \mathcal{T}, \tsopen]; \sigd)$
				\item \quad $(\_, \_, S_{\atte}, S_{\merk}) := \storage[\sid]$; abort if not found 
				\item \quad abort if the $\atte_T^{\id}$ corresponding to $\atte_T^{o}$ is not in $S_{\atte}$
				\item \quad assert $\tsopen$ is within a bounded range with \fnsb.$\xbhb()$
				\item \quad compute 
					$\atte_T^{\od} := \cert([\wtt, \open, \sid, \mt, \tsopen]; \sigd, \sigv)$ 
				\item \quad call \pbc.$\exec(\wtt)$ to trigger on-chain execution 
				\item \quad call \fdapp.$\xopdtb(\atte_T^{\od})$ to acknowledge request 
				\item \quad call \fnsb.$\xaddab(\atte_T^{\od})$ to prove $\atte_T^{\od}$ is sent 
				\item \quad update $S_{\atte}.\add(\atte_T^{\od})$ and $\mt.\status := \opend$
				\item
				\noindent \adjustbox{bgcolor=gray!20,minipage=[t]{\linewidth}}{
					\quad non-blocking wait until \fnsb.$\xmkpb(\atte_T^{\od})$ 
						returns $\merk_T^{\od}$ 
					\item \quad update $S_{\merk}.\add(\merk_T^{\od})$ 
				}
				
				\item \textbf{Upon Receive} \opdt($\atte_T^{\od}$) \public: 
					\hspace*{\fill} \textcolor{olive}{\emph{Southbound}}
				\item \quad ast. $\atte_T^{\od}$ has valid form of  
					$\cert([\wtt, \open, \sid, \mt, \tsopen]; \sigv, \sigd)$  
				\item \quad $(\_, \_, S_{\atte}, \_) := \storage[\sid]$; abort if not found 
				\item \quad abort if the $\atte_T^{o}$ corresponding to 
					$\atte_T^{\od}$ is not in $S_{\atte}$ 
				\item \quad update $S_{\atte}.\add(\atte_T^{\od})$ and $\mt.\status := \opend$
				\vspace*{0.5ex}
				
				\item \textbf{Upon Receive} \clt$(C_\close^\mt)$ \public:  
					\hspace*{\fill} \textcolor{olive}{\emph{Bidirectional}}
				\item \quad assert $C_\close^\mt$ has valid form of  
					$\cert([\wtt, \close, \sid, \mt, \tsclose], \sigd)$ 
				\item \quad assert $\wtt$ is finalized on its destination blockchain and obtain $\merk_\mt^{c_1}$
				\item \quad assert $\tsclose$ is within a bounded margin with \fnsb.$\xbhb()$
				\item \quad $(\_, \_, S_{\atte}, S_{\merk}) := \storage[\sid]$; abort if not found 
				\item \quad compute 
					$\atte_\mt^c := \cert([\wtt, \close, \sid, \mt, \tsclose], \sigd, \sigv)$ 
				\item \quad call \fdapp.$\xcldtb(\atte_\mt^c)$ to acknowledged request 
				\item \quad update $S_{\atte}.\add(\atte_T^c)$, $S_{\merk}.\add(\merk_\mt^{c_1})$ and $\mt.\status := \close$ 
				
				\item \textbf{Upon Receive} \cldt$(\atte_T^c)$ \public: 
					\hspace*{\fill} \textcolor{olive}{\emph{Bidirectional}}
				\item \quad ast. $\atte_T^c$ has valid form of 
					$\cert([\wtt, \close, \sid, \mt, \tsclose], \sigv, \sigd)$ 
				\item \quad $(\_, \_, S_{\atte}, \_) := \storage[\sid]$; abort if not found 
				\item \quad abort if $\cert([\wtt, \close, \sid, \mt, \tsclose], \sigv)$ is not in $S_{\atte}$ 
				\item \quad update $S_{\atte}.\add(\atte_T^c)$ and $\mt.\status := \close$ 
				
				\item \Daemon \redem(\sid) \private:
				\item \quad {\color{comment} \# Invoke the insurance contract periodically}
				\item \quad $(\rgt, \cid, S_{\atte}, S_{\merk}) := \storage[\sid]$; abort if not found 
				\item \quad \textbf{for each} \emph{unclaimed} $\mt \in \rgt$: 
				\item \quad \quad get the $\atte_\mt$ from 
					$S_{\atte}  \bigcup S_{\merk}$ with the most advanced state 
				\item \quad \quad call \fisc.$\xinsub(\cid, \atte_\mt)$ to claim insurance
			\end{enumerate}
		\end{multicols}
		\vspace*{-0.05in}
		}
	}
        }
	\caption{Protocol description of of \fven.  
		\protect\adjustbox{bgcolor=gray!20}{Gray background} 
		denotes non-blocking operations triggered by
		status updates on \fnsb. Handlers annotated with \emph{northbound}
		and \emph{southbound} process transactions originated from \fven and \fdapp, respectively. 
	Handlers annotated with \emph{bidirectional} are shared by all transactions.}
	\label{fig:fven}
\end{figure*}
\subsection{Execution Protocol by \VENs}\label{sec:fven}
The full protocol of \fven is detailed in \cref{fig:fven}. Below we clarify 
some technical subtleties. 

\subsubsection{Post Compilation and Session Setup}
After \gt is generated, \fven initiates an execution session for \gt in the \xpcmplb daemon. 
The primary goal of the initialization is to create and deploy an insurance 
contract to protect the execution of \gt. Towards this end, 
\fven interacts with the protocol \fisc to create the insurance \contract 
for \gt, and further deploys the \contract on \NSB after the \dApp client $\md$ 
agrees on the \contract. Throughout the paper, 
$\cert([*]; \sig)$ represents a signed certificate 
proving that the signing party agrees on the value enclosed in 
the certificate. We use $\sigv$ and $\sigd$ to represent the 
signature by \fven and \fdapp, respectively. 

Additionally, both \fven and \fdapp are required to deposit sufficient 
funds to \fisc to ensure that \fisc holds sufficient funds 
to financially revert all committed transactions 
regardless of the step at which the execution aborts prematurely. 
Intuitively, each party would need to stake 
at least the total amount of incoming funds to the 
party \emph{without} deducting the outgoing funds. 
This strawman design, however, require 
high stakes. More desirably, considering the dependency requirements in \gt, 
an party $\mx$ (\fven or \fdapp) only needs to stake 
$$\max_{s\in \mathcal{G}_S} \sum_{\mt\in s ~\wedge~ \mt.\textsf{to} = \mx} \mt.\hamt.\amt -
\sum_{\mt\in s ~\wedge~ \mt.\textsf{from} = \mx} \mt.\hamt.\amt$$
where $\mathcal{G}_S$ is the set of all committable subsets in 
\gt, where a subset $s\subseteq\mathcal{G}_T$ is \emph{committable}
if, whenever $\mt\in s$, all preconditions of $\mt$ are also in $s$.
For clarity of notation, throughout the paper, when 
saying $\mt.\textsf{from} = $\fven or $\mt$ is originated from \fven, 
we mean that $\mt$ is sent and signed by an account owned by \fven. Likewise, 
$\mt.\textsf{from} = $\fdapp indicates that $\mt$ is sent from an account entity 
defined in the \HSL program. 
\fisc refunds any remaining funds after the contract is terminated.  

After the \contract is instantiated and sufficiently staked, 
\fven initializes its internal bookkeeping for the session. 
The two notations $S_{\atte}$ and $S_{\merk}$ represent two sets that store the 
signed certificates received via off-chain channels and on-chain 
Merkle proofs constructed using \fnsb and \pbc. 

\subsubsection{Protocol Exchange for Transaction Handling}\label{sec:fven:exchange}
In \fven, \xsditedb and \xoptb are two handlers processing 
\emph{northbound} transactions which originates from \fven. 
The \xsditedb handling for $\mt$ is invoked when all 
its preconditions are finalized, which is detected 
by the watching service of \fven (\cf \cref{sec:fven:watching}). 
The \xsditedb computes $\atte_\mt^{\id}$ to prove $\mt$ is in  
the \initd state , and then passes it to the corresponding handler of \fdapp 
for subsequent processing. Meanwhile, \xsditedb  
stakes $\atte_\mt^{\id}$ on \fnsb, and later it retrieves a Merkle 
proof $\merk_\mt^i$ from the \NSB to prove that $\atte_\mt^{\id}$ has been sent. 
$\merk_\mt^{\id}$ essentially is a hash chain linking $\atte_\mt^{\id}$ back to 
an \textsf{ActionRoot} on a committed block of the \NSB. 
The proof retrieval is a non-blocking operation triggered by the consensus update on the \NSB. 
 
The \xoptb handler pairs with \xsditedb. It listens for a 
timestamped $\atte_\mt^o$, which is supposed to be generated by 
\fdapp after it processes $\atte_\mt^{\id}$ from \fven. 
\xoptb performs special correctness check on the $\tsopen$ enclosed in $\atte_\mt^o$. 
In particular, \fven and \fdapp use the block height of the \NSB as a calibrated clock. 
By checking that $\tsopen$ is within a bounded range of the \NSB height, 
\fven ensures that the $\tsopen$ added by \fdapp is fresh. 
After all correctness checks on $\atte_\mt^{\id}$ are passed, the state of $\mt$ 
is promoted from \open to \opend. \xoptb then computes   
certificate to prove the updated state and posts $\wtt$ on 
its destination blockchain for on-chain execution. 
Throughout the paper, $\wtt$ denotes the on-chain executable transaction 
computed and signed using the information contained in $\mt$. 
Note that the difference between the $\atte_\mt^o$ received from \fdapp and a 
post-open (\ie \opend) certificate $\atte_\mt^{\od}$ 
computed by \fven is that latter one is 
signed by both parties. Only the $\tsopen$ specified in $\atte_\mt^{\od}$ 
is used by \fisc when evaluating the deadline constraint of $\mt$. 

Southbound transactions originating from \fdapp are processed by 
\fven in a similar manner as the northbound transactions, via  
the \xrvitedb and \xopdtb handlers. We clarify a subtlety in the \xrvitedb handler when  
verifying the \emph{association} between $\wtt$ and $\mt$ (line \ref{line:associate}). 
If $\wtt$ depends on the resulting state from its upstream transactions (for instance, 
\emph{T4} depends on the resulting state of \emph{T1} in \cref{fig:tdg}), 
\fven needs to verify that the state used by $\wtt$ is consistent with 
the state enclosed in the finalization proofs of those upstream transactions. 

\begin{figure}[t]
	\centering
	\mbox{
		\includegraphics[width=0.95\columnwidth]{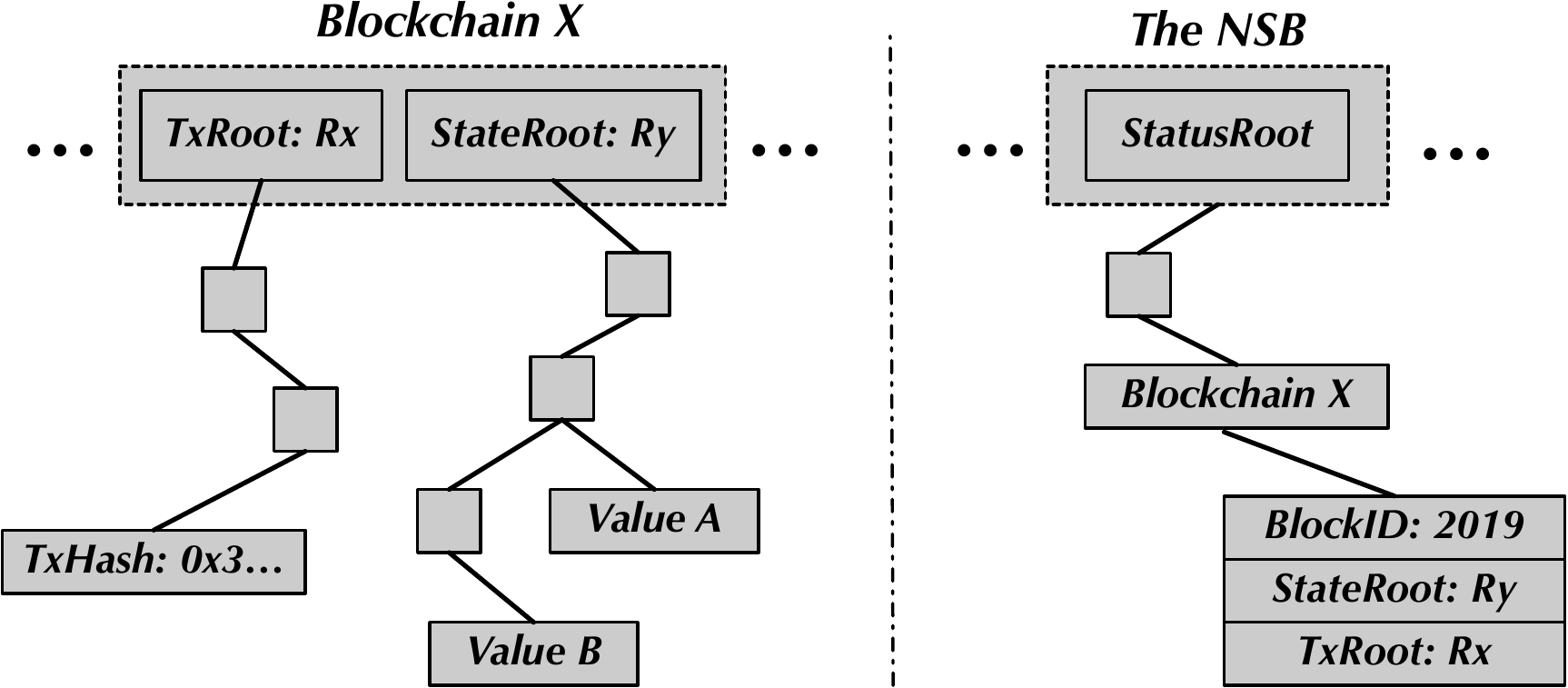}
	}
	\caption{The complete on-chain proof (denoted by $\merk_\mt^{c}$) to prove that the state of a transaction 
	is eligible to be promoted as \close. The left-side part is the finalization proof (denoted by $\merk_\mt^{c_1}$) for the transaction collected from its 
	destination blockchain; the right-side part is the blockchain status proof (denoted by $\merk_\mt^{c_2}$) collected from the \NSB.} \label{fig:merkle_proof}
\end{figure}

\newcommand{\fund}{\textsf{F}_{\textsf{stake}}\xspace}
\newcommand{\rst}{\textsf{st}_{\textsf{result}}\xspace}
\newcommand{\stproof}{\textsf{st}_{\textsf{proof}}\xspace}
\newcommand{\tr}{\textsf{T}}
\newcommand{\mpvf}{\textsf{MerkleVerify}}
\newcommand{\livevf}{\textsf{DeadlineVerify}}
\newcommand{\sigvf}{\textsf{SigVerify}}
\newcommand{\dtrans}{\textsf{DirtyTrans}}

\begin{figure*}[t]
	\small
	\fbox {
		\parbox{\textwidth}{
			\vspace*{-0.15in}
			\begin{multicols}{2}
				\begin{enumerate}[label=\textbf{\scriptsize \arabic*}, leftmargin=*,itemsep=0.25ex]
					\item \Init $\storage := \emptyset$ 
					
					\item \textbf{Upon Receive} $\xcrc(\rgt)$:
					\item \quad generate the arbitration cod, denoted by \contract, as follows 
					\item \quad initialize three maps $\st$, $\Ag$ and $\fund$ 
					\item \quad \textbf{for each} $\mt \in \rgt$ \bco 
					\item \quad \quad compute an internal identifier for $\mt$ as $\tid := H(\mathcal{T})$ 
					\item \quad \quad initialize $\st[\tid] := [\unknown, \mt, \tsopen{=}0, \tsclose{=}0, \stproof]$ 
					\item \quad \quad retrieve \tid's fund-reversion account, denoted as \dst 
					\item \quad \quad initialize $\Ag[\tid] := [\amt{=}0, \dst]$ 
					\item \quad compute an identifier for \contract as
					$\cid := H(\overrightarrow{0}, \contract)$ 
					\item \quad initialize $\storage[\cid] := [\rgt, \st, \Ag, \fund]$ 
					\item \quad send $[\cid, \contract]$ to the requester for acknowledgment
					
					\item \textbf{Upon Receive} $\xskf(\cid)$: 
					\item \quad $(\_, \_, \_, \_, \fund) := \storage[\cid]$; abort if not found 
					\item \quad update $\fund[\textsf{msg}.\textsf{sender}] := 
					\fund[\textsf{msg}.\textsf{sender}] + \textsf{msg}.\textsf{value}$ 
					
					\item \textbf{Upon Receive} \insu($\cid, \realatte$): 
					\item \quad $(\_, \_, \st, \_, \_) := \storage[\cid]$; abort if not found 
					\item \quad compute $\tid := H(\realatte.\mt)$; $\tr := \st[\tid]$ abort if not found
					\item \quad abort if $\tr.\status$ is \emph{more advanced} the state enclosed by $\atte$ 
					\item \quad \textbf{if} $\realatte$ is a certificate signed by both parties \bco 
					\item \quad \quad assert $\sigvf(\realatte)$ is true 
					\item \quad \quad \textbf{if} $\realatte$ is $\atte_\mt^{\od}$ \bco
					update $\tr.\status := \opend$; $\tr.\tsopen := \realatte.\tsopen$ 
					\item \quad \quad \textbf{else} \bco update $\tr.\status := \close$; $\tr.\tsclose := \realatte.\tsclose$ 
					\item \quad \textbf{else} \bco~ {\color{comment} \# \realatte is in form of a Merkle proof} 
					\item \quad \quad assert $\mpvf(\realatte)$ is true 
					\item \quad \quad \textbf{if} 
					$\realatte$ is a $\merk_\mt^{i}$ \textbf{or} 
					$\merk_\mt^{\id}$ \textbf{or} 
					$\merk_\mt^{o}$ \bco 
					\item \quad \quad \quad retrieve the certificate  
					$\atte_\mt^{i}$ or $\atte_\mt^{\id}$ or $\atte_\mt^{o}$ from \realatte 
					\item \quad \quad \quad 
					assert the $\wtt$ enclosed in $\atte_\mt^{\id}$ or $\atte_\mt^o$ is genuine 
					\item \quad \quad \quad 
					assert the $\tsopen$ enclosed in $\atte_\mt^o$ is genuine 
					\item \quad \quad \quad update $\tr.\status := \realatte.\status$ 
					\item \quad \quad \textbf{elif} $\realatte$ is $\merk_\mt^{\od}$ \bco 
					\item \quad \quad \quad retrieve the certificate $\atte_\mt^{\od}$ from \realatte 
					\item \quad \quad \quad 
					update $\tr.\status := \opend$ and $\tr.\tsopen :=  \atte_\mt^{\od}.\tsopen$ 
					\item \quad \quad \textbf{elif} $\realatte$ is $\merk_\mt^{c}$ \bco 
					\item \quad \quad \quad 
					update $\tr.\stproof$ based on $\merk_\mt^{c_1}$ if necessary 
					\item \quad \quad \quad 
					update $\tr.\tsclose$ as the height of the block attaching $\merk_\mt^{c_2}$ 
					\item \quad \quad \quad update $\tr.\status := \close$  
					
					\item \textbf{Upon Timeout} \setc(\cid):
					\hspace*{\fill} \textcolor{olive}{\emph{Internal Daemon}}
					\item \quad $(\rgt, \st, \Ag, \fund) := \storage[\cid]$; abort if not found 
					\item \quad \textbf{for} $(\tid, \tr) \in \st$ \bco 
					\item \quad \quad \textbf{continue} if $\tr.\status$ is not \close 
					\item \quad \quad update $\Ag[\tid].\amt := \tr.\mt.\hamt.\amt$ 
					\item \quad \quad \textbf{if} $\livevf(\tr) = \textbf{true}$ \bco update $\tr.\status := \cort$ 
					\item \quad compute $\ms := \dtrans(\rgt, \st)$ {\color{comment} 
						\# non-empty if execution fails.}
					\item \quad execute fund reversion for non-zero entries 
					in $\Ag$ if $\ms$ is not empty
					\item \quad initialize a map \resp to record which party to blame 
					\item \quad \textbf{for each}  $(\tid, \tr) \in \ms$ \bco  
					\item \quad \quad \textbf{if} 
					$\tr.\status = \close ~|~ \open ~|~ \opend$ \bco 
					$\resp[\tid] := \tr.\mt.\from$ 
					\item \quad \quad \textbf{elif} $\tr.\status = \initd$ \bco 
					$\resp[\tid] := \tr.\mt.\textsf{to}$
					\item \quad \quad \textbf{elif} $\tr.\status = \init$ \bco 
					$\resp[\tid] := \md$
					\item \quad \quad \textbf{else} \bco $\resp[\tid] := \mv$
					\item \quad return any remaining funds in $\fund$ to corresponding senders 
					\item \quad call $\storage.\textsf{erase}[\cid]$ to stay silent afterwards
				\end{enumerate}
			\end{multicols}
			\vspace*{-0.05in}
		}
	}
	\caption{\fisc: the protocol realization of the \ISC arbitrator.}\label{fig:fisc}
\end{figure*}

\subsubsection{Proactive Watching Services}\label{sec:fven:watching}
The cross-chain execution process proceeds when all 
session-relevant blockchains and the \NSB make progress on transactions. 
As the driver of execution, \fven internally creates two watching services to 
\emph{proactively} read the status of those blockchains. 
 
In the watching daemon to one blockchain, \fven mainly reads the public ledger 
of \pbc to monitor the status of transactions that have been posted 
for on-chain execution. If \fven notices that an on-chain transaction 
$\wtt$ is recently finalized, it requests the closing process for $\mt$ by sending \fdapp a 
timestamped certificate $C_\close$. The pair of handlers, \xcltb and \xcldtb, are 
used by both \fven and \fdapp in this exchange. 
Both handlers can be used for handling northbound and southbound 
transactions, depending on which party sends the closing request. 
In general, a transaction's originator 
has a stronger motivation to initiate the closing process  
because the originator would be held accountable 
if the transaction were not timely closed by its deadline.

In addition, \fven needs to retrieve a Merkle Proof from \pbc to prove the finalization of $\wtt$. 
This proof, denoted by $\merk_\mt^{c_1}$, serves two purposes: 
\first it is the first part of a complete on-chain proof to prove that 
the state $\wtt$ can be promoted to \close, as shown in \cref{fig:merkle_proof}; 
\second if the resulting state of $\wtt$ is used by its downstream transactions, 
$\merk_\mt^{c_1}$ is necessary to ensure that those downstream transactions indeed 
use genuine state. 

In the watching service to \fnsb, \fven performs following tasks. First, 
as described in \cref{sec:fven:channel}, \NSB is treated as a fallback communication 
medium for the off-chain channel. Thus, \fven searches the sorted \acmt to look for any session-relevant 
certificates that have not been received via the off-chain channel.
Second, for each \opend $\mt$ whose \close attestation 
is still missing after \fven has sent $C_\close$ (indicating slow or no reaction from \fdapp), 
\fven tries to retrieve the second part of $\merk_\mt^{c}$ from \fnsb. 
The second proof, denoted as $\merk_\mt^{c_2}$, is to prove that the Merkle roots referred in $\merk_\mt^{c_1}$ 
are correctly linked to a \textsf{StatusRoot} on a finalized \NSB block (see \cref{fig:merkle_proof}).  
Once $\merk_\mt^{c}$ is fully constructed, the state of $\mt$ is promoted as \close. 
Finally, \fven may find a new set of transactions that are eligible to be executed 
if their preconditions are finalized due to any recently-closed transactions. If so, \fven processes 
them by either requesting initialization from \fdapp or 
calling \xsditedb internally, depending on the originators of those transactions. 

\subsubsection{\fisc Invocation} 
\fven periodically invokes \fisc to execute the contract. All internally 
stored certificates and \emph{complete} Merkle proofs are acceptable. 
However, for any $\mt$, \fven should invoke \fisc only 
using the attestation with the most advanced state, since lower-ranked attestations for $\mt$ 
are effectively ignored by \fisc (\cf \S~\ref{sec:UIP:fisc}).

\vspace{-0.2cm}
\subsection{Execution Protocol by \dApp Clients}\label{sec:fdapp}
\fdapp specifies the protocol implemented by \dApp clients. 
\fdapp defines the following set of handlers to match \fven. 
In particular, the \xitdtb and \xopdtb match the \xsditedb and \xoptb of \fven, respectively, 
to process $\atte^{\id}$ and $\atte^{\od}$ sent by \fven 
when handling transactions originated from \fven. The \xinittb and \xoptb process 
$\atte^i$ and $\atte^o$ sent by \fven when executing transactions originated from \fdapp. 
The \xcltb and \xcldtb of \fdapp match their counterparts in \fven to negotiate  
closing attestations. 

For usability, \sys imposes smaller requirements on the watching daemons implemented by \fdapp. Specially, 
\fdapp still proactively watches \fnsb to have a fallback communication medium with \fven. 
However, \fdapp is \emph{not} required to proactively watch the 
status of underlying blockchains or dynamically compute eligible 
transactions whenever the execution status changes. 
We intentionally offload such complexity on \fven to 
enable lightweight \dApp clients. \fdapp, though, should (and is motivated to) check the status of 
self-originated transactions in order to request transaction closing. 
\vspace{-0.1cm}
\subsection{Protocol Realization of the \ISC}\label{sec:UIP:fisc}
\cref{fig:fisc} specifies the protocol realization of the \ISC. 
The \xcrcb handler is the entry point of requesting insurance contract creation 
using \fisc. It generates the arbitration code, denoted as \contract, based on the given \dApp executable \gt. 
The \contract internally uses $\st$ to track the state of each transaction in \gt, which is 
updated when processing security attestations in the \xinsub handler. 
For clear presentation, \cref{fig:fisc} extracts the state proof and fund reversion tuple from $\mt$ as 
dedicated variables $\stproof$ and $\Ag$. 
When the \fisc times out, it executes the contract terms based on its internal state, after which its
funds are depleted and the contract never runs again. Below we explain several technical subtleties. 

\subsubsection{Insurance Claim}\label{sec:fisc:claim}
The \xinsub handler processes security attestations from \fven and \fdapp. Only dual-signed certificates  
(\ie $\atte^{\od}$ and $\atte^{c}$) or complete Merkle proofs are acceptable. Processing 
dual-signed certificates is straightforward as they are explicitly agreed by both parties. 
However, processing Merkle proof requires additional correctness checks. First, when validating a Merkle 
proof $\merk_\mt^{i}$,  $\merk_\mt^{\id}$ or $\merk_\mt^{o}$, 
\fisc retrieves the single-party signed certificate $\atte_\mt^{i}$, $\atte_\mt^{\id}$ or $\atte_\mt^{o}$ 
enclosed in the proof and performs the following correctness check against the certificate. \first 
The certificate must be signed by the correct party, \ie $\atte_\mt^{i}$ is signed by \fven, 
$\atte_\mt^{\id}$ is signed by $\mt$'s originator and $\atte_\mt^{o}$ is signed by the destination of 
$\mt$. \second The enclosed on-chain transaction $\wtt$ in $\atte_\mt^{\id}$ and $\atte_\mt^o$ is correctly associated 
with $\mt$. The checking logic is the same as the on used by \fven, which has been 
explained in \cref{sec:fven:exchange}. \third The enclosed $\tsopen$ in $\atte_\mt^o$ is genuine, 
where the genuineness is defined as a bounded difference between $\tsopen$ and 
the height of the \NSB block that attaches $\merk_\mt^o$. 

\begin{figure}[t]
	\centering
	\mbox{
		\includegraphics[width=\columnwidth]{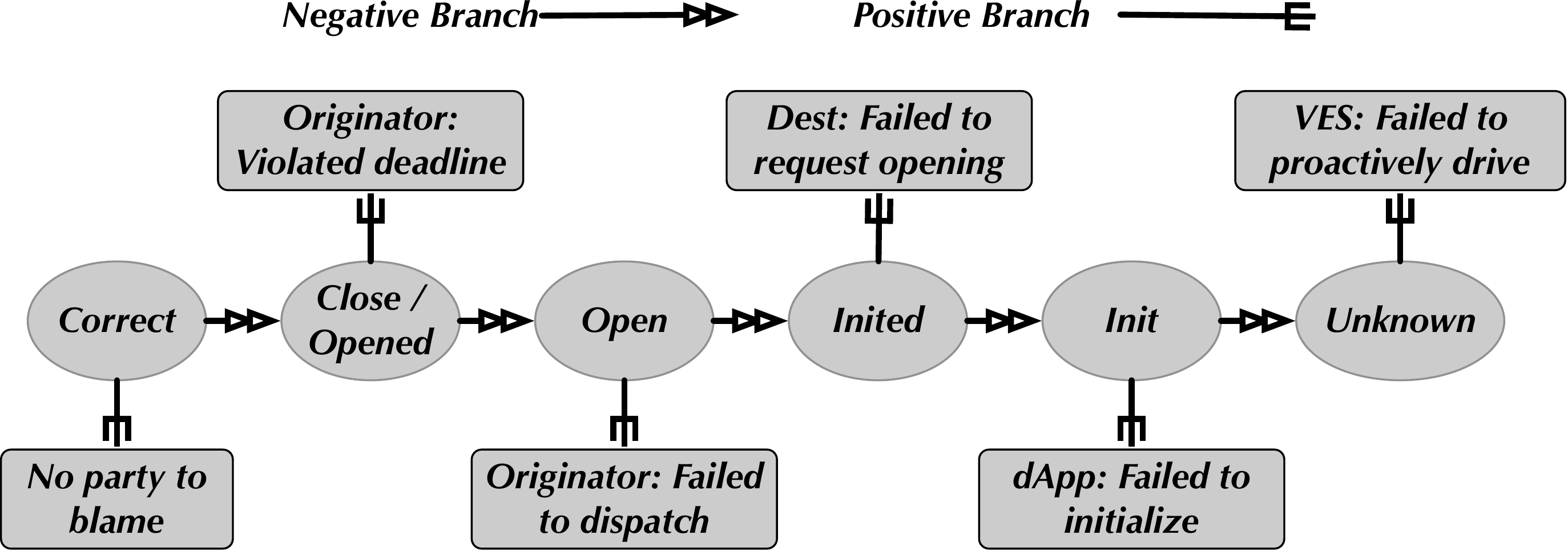}
	}
	\caption{The decision tree to decide the accountable party for a dirty 
		transaction.} \label{fig:decision}
\end{figure}

\subsubsection{Contract Term Settlement}
\fisc registers a callback \xsetcb to execute contract terms automatically upon 
timeout. \fisc internally defines an additional transaction state, called \cort. The state of a \close transaction is promoted to \cort if 
its deadline constraint is satisfied. Then, \fisc computes the  possible \emph{dirty} transactions in \gt, which are the 
transactions that are eligible to be opened, but with non-\cort 
state. Thus, the execution succeeds only if \gt has no dirty transactions. 
Otherwise, \fisc employs a decision tree, shown in Figure~\ref{fig:decision}, 
to decide the responsible party for each dirty transaction. 
The decision tree is derived from the execution steps taken by \fven and 
\fdapp. In particular, if a transaction $\mt$'s state is \close, 
\opend or \open, then it is $\mt$'s originator to blame for 
either failing to fulfill the deadline constraint or failing to 
dispatch $\wtt$ for on-chain execution. If a transaction $\mt$'s 
state is \initd, then it is $\mt$'s destination party's 
responsibility for not proceeding with $\mt$ even though 
$\atte_\mt^{\id}$ has been provably sent. If a transaction $\mt$'s 
state is \init (only transactions originated from \dApp $\md$ can 
have \init status), then $\md$ (the originator) is the party to blame for 
not reacting on the $\atte_\mt^i$ sent by $\mv$. 
Finally, if transaction $\mt$'s state is \unknown, then 
$\mv$ is held accountable for not proactively driving the 
initialization of $\mt$, no matter which party originates $\mt$.

\hyphenation{block-chain}
\hyphenation{block-chains}
\subsection{Specification of \fnsb and \pbc}\label{sec:UIP:fnsb}
\pbc specifies the protocol realization of a general-purpose 
blockchain where a set of consensus nodes run a 
secure protocol to agree upon the 
public global state. In this paper, we regard \pbc 
as a conceptual party trusted for correctness 
and availability, \ie \pbc guarantees to correctly 
perform any predefined computation (\eg Turing-complete 
smart contract programs) and is always available to 
handle user requests despite unbounded response 
latency. \fnsb specifies the protocol realization of the \NSB. 
\fnsb is an extended version of \pbc with additional capabilities. Due to space constraint, 
we move the detailed protocol description of \pbc and \fnsb to \cref{sec:appendix:fnsb_spec}.

\subsection{Security Theorems}\label{sec:UIP:theory}
To rigorously prove the security properties of \UIP, we first present the
cryptography abstraction of the \UIP in form of an ideal functionality
\fbip. The ideal functionality articulates the correctness and security
properties that \UIP wishes to attain by assuming a trusted entity.
Then we prove that \pbip, our the decentralized real-world 
protocol containing the aforementioned preliminary protocols, 
securely realizes \fbip using the UC framework~\cite{UC}, \ie \pbip achieves the same functionality
and security properties as \fbip without assuming any trusted authorities. 
Since the rigorous proof requires non-trivial simulator construction within the UC framework, 
we defer detailed proof to a dedicated section \cref{sec:appendix:the_prof}. 
\section{Implementation and Experiments}\label{sec:imple}
In this section, we present the implementation of a \sys prototype and 
report experiment results on the prototype. 
At the time of writing, the total development effort includes 
\first ${\sim}$1,500 lines of Java code and ${\sim}$3,100 lines of ANTLR~\cite{antlr} grammar code 
for building the \HSL programming framework, \second 
${\sim}$21,000 lines of code, mainly in Go and Python, for implementing the \UIP protocol; and  
${\sim}$8,000 lines of code, mainly in Go, for implementing the \NSB; 
and \third ${\sim}$1,000 lines of code, in Solidity, Vyper, Go and \HSL, for writing cross-chain \dApps 
running on \sys. The released source code is available at~\cite{source}. 
The \sys Consortium is under active code maintenance and new feature 
development for \sys. 

\subsection{Platform Implementation}\label{sec:imple:platform}
To demonstrate the interoperability and programmability across heterogeneous blockchains 
on \sys, our current prototype incorporates Ethereum, the flagship public blockchain, 
and a permissioned blockchain built atop the Tendermint~\cite{tendermint} consensus 
engine, a commonly cited cornerstone for building enterprise blockchains. 
We implement the necessary accounts (wallets), the smart contract environment, 
and the on-chain storage to deliver the permissioned blockchain with full programmability. 
The \NSB is also built atop Tendermint with full 
support for its claimed capabilities, such as action staking and Merkle proof retrieval.

For the programming framework, we implement the \HSL compiler that takes \HSL programs and 
contracts written in Solidity, Vyper, and Go as input, and produces transaction dependency graphs.
We implement the multi-lang front end and the \HSL front end using ANTLR~\cite{antlr},
which parse the input \HSL program and contracts, build an intermediate representation of the \HSL program,
and convert the types of contract entities into our unified types.
We also implement the validation component that analyzes the intermediate representation 
to validate the entities, operations, and dependencies specified in the \HSL program.

Our experience with the prototype implementation is that \emph{the 
effort for horizontally scaling \sys to incorporate a new blockchain is lightweight}: 
it requires no protocol change to both \UIP and the blockchain itself. 
We simply need to add an extra parser to the multi-lang front end to support 
the programming language used by the blockchain (if this language has not been supported by \sys), and 
meanwhile \VENs extend their visibility to this blockchain. 
The \sys consortium is continuously working on on-boarding additional 
blockchains, both permissioned and permissionless. 

\subsection{Application Implementation}\label{sec:imple:app}
Besides the platform implementation, we further implement and deploy three categories of 
cross-chain \dApps on \sys. 

\parab{Financial Derivatives.} Financial derivatives are among the mostly cited blockchain applications. 
However, external data feed, \ie an oracle, is often required for financial instructions.
Currently, oracles are either built atop trusted third-party providers (\eg Oraclize~\cite{oraclize}),
or using trusted hardware enclaves~\cite{TC}. \sys, for the first time, realizes
the possibility of \emph{using blockchains themselves as oracles}. With the built-in
decentralization and correctness guarantees of blockchains, \sys 
fully avoids trusted parties while delivering genuine data feed to
smart contracts. In this application sector, we implement a cross-chain cash-settled Option 
\dApp in which options can be natively traded on different blockchains (a scaled-up version 
of the introductory example in \cref{sec:overview:hsl}). 

\parab{Cross-Chain Asset Movement.} \sys natively enables cross-chain 
asset transfers without relying on any trusted entities, such as exchanges. 
This primitive could power a wide range of applications, such as a global 
payment network that interconnects geographically distributed bank-backed consortium
blockchains~\cite{chase}, an initial coin offering in which tokens can 
be sold in various cryptocurrencies, and a gaming platform where players 
can freely trade and redeem their valuables (in form of non-fungible tokens) across different games. 
In this category, we implement an asset movement \dApp with hybrid operations 
where assets are moved among accounts and smart contracts across different 
blockchains

\parab{Federated Computing.} In a federated computing model, all 
participants collectively work on an umbrella task by submitting their local 
computation results. In the scenario where transparency and accountability 
are desired, blockchains are perfect platforms for persisting both 
the results submitted by each participant and the logic for aggregating 
those results. In this application category, we implement a federated 
voting system where delegates in different regions can submit their votes to 
their regional blockchains, and the logic for computing the final votes based on the
 regional votes is publicly visible on another blockchain. 
\subsection{Experiments}\label{sec:evaluation}
We ran experiments with three blockchain testnets: one private Ethereum testnet, one
Tendermint-based blockchain, and the \NSB. 
Each of those testnets is deployed on a VM instance of a public cloud on different continents. 
For experiment purposes, \dApp clients and \VEN nodes can be deployed either locally or on cloud. 

\begin{table}[t]
	\resizebox{\columnwidth}{!}{%
		\begin{tabular}{c||c|c||c|c||c|c}
			\hline
			\multirow{2}{*}{} & \multicolumn{2}{c|}{\begin{tabular}[c]{@{}c@{}}Financial \\ Derivatives\end{tabular}} & \multicolumn{2}{c|}{\begin{tabular}[c]{@{}c@{}}CryptoAsset\\ Movement\end{tabular}} & \multicolumn{2}{c}{\begin{tabular}[c]{@{}c@{}}Federated \\ Computing\end{tabular}} \\ \cline{2-7} 
				& Mean & \% & Mean & \% & Mean & \% \\ \hline\hline
				HSL Compilation & 1.1769 & $\sim$16 & 0.2598 & $\sim$4 & 1.095 & $\sim$15 \\ \hline
				Session Creation & 4.2399 & $\sim$58 & 4.1529 & $\sim$67 & 4.2058 &  $\sim$60 \\ \hline
				Action/Status Staking & 0.6754 & $\sim$10 & 0.7295 & $\sim$12 &  0.7592 & $\sim$11 \\ \hline
				Proof Retrieval & 1.0472 & $\sim$15 & 1.0511 & $\sim$17 & 0.9875 & $\sim$14 \\ \hline\hline
				Total & 7.1104 &  & 6.1933 &   & 7.0475 &  \\ \hline
			\end{tabular}
		}
		\caption{End-to-end \dApp execution latency on \sys, with profiling breakdown. All times are in seconds.}
		\label{tab:e2e}
	\end{table}

\subsubsection{End-to-End Latency}\label{sec:eval:e2e}
We evaluated all three applications mentioned in \cref{sec:imple:app} and 
reported their end-to-end execution 
latency introduced by \sys in \cref{tab:e2e}. The reported latency 
includes \HSL program compiling, \dApp-\VEN session 
creation, and (batched) \NSB action staking and proof retrieval during the \UIP protocol exchange. 
All reported times include the networking latency across the global Internet.
Each datapoint is the average of more than one hundred runs. 
We do not include the latency for actual on-chain execution since the consensus 
efficiency of different blockchains varies and is not controlled by \sys. We also do not include 
the time for \ISC insurance claims in the end-to-end latency because they can be 
done offline anytime before the \ISC expires. 

These \dApps show similar latency profiling breakdown, where 
the session creation is the most time consuming phase because it requires 
handshakes between the \dApp client and \VEN, and also includes 
the time for \ISC deployment and initialization. The CryptoAsset \dApp has a much lower 
\HSL compilation latency since its operation only involves one smart contract, 
whereas the rest two \dApps import three contracts written in Go, Vyper, and Solidity.
In each \dApp, all its \NSB-related operations (\eg action/status staking and proof retrievals) 
are bundled and performed in a batch for experiment purpose, 
even though all certificates required for \ISC arbitration have been received via off-chain channels. 
The sizes of actions and proofs for three \dApps are different since their executables contain  
different number of transactions. 

\begin{figure}[t]
	\centering
	\mbox{
		\includegraphics[width=0.95\columnwidth]{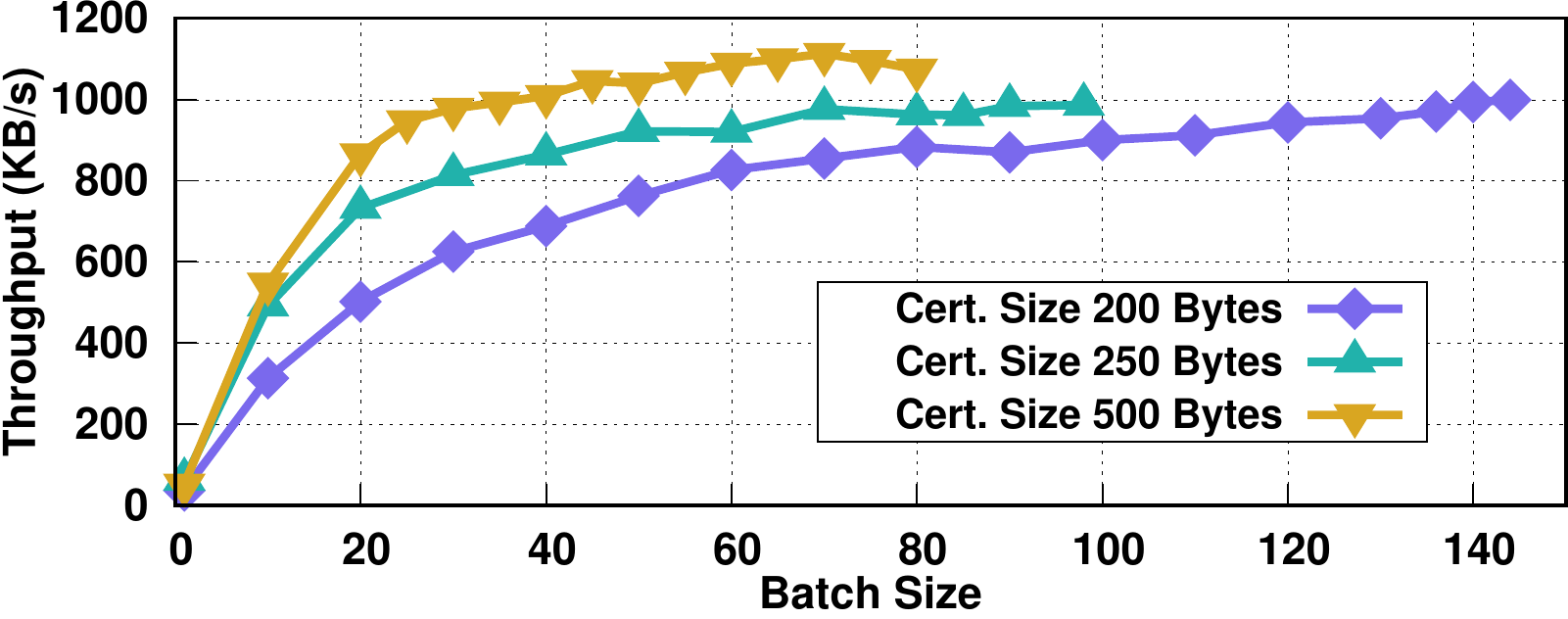}
	}
	\caption{The throughput of the \NSB, measured as the total size of 
		committed certificates on the \NSB per second.} \label{fig:throughput}
\end{figure}

\subsubsection{\NSB Throughput and \sys Capacity}\label{sec:eval:thro}
The throughput of the \NSB affects aggregated \dApp 
capacity on \sys. In this section, we report the peak throughput of the currently implemented \NSB. 
We stress tested the \NSB by initiating up to one thousand \dApp clients and \VEN nodes, which 
concurrently dispatched action and status staking to the \NSB. 
We batched multiple certificate stakings by different clients into a single \NSB-transaction, so that the effective 
certificate-staking throughput perceived by those clients can exceed the consensus limit of the \NSB.  
Figure~\ref{fig:throughput} plots the \NSB throughput, measured as the total size of committed certificates 
by all clients per second, under different certificate and batch sizes. The results show that as the batch 
size increases, regardless of the certificate sizes, 
the \NSB throughput converged to about $1000$ kilobytes per second. 
Given any certificate size, further enlarging the batch size cannot boost the throughput, whereas 
the failure rate of certificate staking increases, indicating that the \NSB is fully loaded. 

Given the above \NSB throughput, the actual \dApp capacity of the \sys platform 
further depends on how often the communication between \dApp clients and \VENs 
falls back to the \NSB. In particular, each \dApp-transaction  
spawns at most six \NSB-transactions (five action stakings and one status staking), 
assuming that the off-chain channel is fully nonfunctional (zero \NSB transaction if otherwise). 
Thus, the \emph{lower bound} of the aggregate \dApp capacity on \sys, 
which would be reached only if all off-chain channels among \dApp clients 
and \VENs were simultaneously broken, is about $\frac{170000}{s}$ 
transactions per second (TPS), where $s$ is the (average) size (in bytes) of a certificate. 
This capacity and the TPS of most PoS production blockchains are of the same magnitude. 
Further, considering \first the \NSB is horizontally shardable  
at the granularity of each underlying blockchain (\cref{sec:uip:nsb}) and 
\second not all transactions on an underlying blockchain are cross-chain related, 
we anticipate that the \NSB will not become the bottleneck 
as \sys scales to support more blockchains in the future.

\section{DISCUSSION}\label{sec:discussion}
In this section, we discuss several aspects that have not been thoroughly addressed 
in this paper, and present our vision for future work on \sys and its impact. 

\subsection{Programming Framework Extension}\label{discussion:hsl}
\HSL is a high-level programming language designed to write cross-chain \dApps under the \USM programming model. 
The language constructs provided by \HSL allow developers to directly specify entities, operations, and dependencies in \HSL programs. 
To ensure the determinism of operations, which is an important property for the \NSB and the \ISC to determine the correctness or violation of \dApp executions, 
the language constructs do not include control-flow operations such as conditional branching, looping, and calling/returning from a procedure.
Additionally, dynamic transaction generation is also not supported by \HSL, 
since it has led to a new class of bugs
known as re-entrancy vulnerabilities~\cite{reentrancy}.
These design choices are consistent with the recent blockchain programming languages that emphasize on safety guarantees, such as Move~\cite{movelang} for Facebook's Libra blockchain.

In future work, we plan to extend the design of the \UIP protocol to 
support \emph{dynamic transaction graphs}, which allows 
conditional execution of operations and certain degree of 
indeterminism of operation executions, such as repeating an operation for a specific times based on the values of state variables computed from previous operations. 
With those extensions, we are able to implement control-flow 
operations into \HSL and provide both static and 
dynamic verification to ensure the correctness of \dApps.

\subsection{Cross-Shards and Cross-Worlds}\label{discussion:cross_shard}
\sys is motivated by heterogeneous blockchain interoperation. 
Thanks to its generic design, \sys can also 
enable cross-shard smart contracting and transactions for   
sharded blockchain platforms (\eg OmniLedger~\cite{omniledger} and RapidChain~\cite{rapidchain}). 
On the one hand, the \HSL programming framework 
is blockchain-neutral and extensible. Thus, writing \dApps that
involve smart contracts and accounts on different blockchains 
is conceptually identical to writing \dApps that operate contracts and accounts on 
different shards. In fact, given that most of those sharded blockchains are 
homogeneously sharded (\ie all shards have the same format of 
contracts and accounts), developing and compiling cross-shard 
\dApps using \HSL are even simpler than cross-chain \dApps. 
On the other hand, realizing \UIP on sharded blockchains also 
requires less overhead since maintaining an \NSB 
for all (homogeneous) shards is more lightweight 
than maintaining an \NSB supporting heterogeneous blockchains.  
In fact, many sharded blockchain platforms already 
maintain a dedicated global blockchain as their trust anchor 
(\eg the identity chain of OmniLedger~\cite{omniledger} and 
the beacon chain of Harmony~\cite{harmony}),
to which the \NSB functionality can be ported.

Additionally, we envision that the fully connected Web 3.0 should also 
include centralized platforms (\ie Cloud) to compensate for 
functionality (\eg performing computationally intensive tasks)  
that is difficult to execute on-chain. 
We recognize that two additional capabilities, with 
minimal distribution of their operation models, are required from those centralized 
platforms to make them compatible with \sys: \first any public state they publish should 
be coupled with verifiable proofs to certify the correctness of the state (where 
the definition of correctness could be application-specific), and \second 
all published state should have the concept of finality. 
With such capabilities, \dApps on \sys can trustlessly 
incorporate the state published by those centralized platforms. 

\subsection{Cross-Chain Interoperability Service Providers}\label{discussion:csp}
\VENs play vital roles on \sys platform. We envision that \VENs would enter the 
\sys ecosystem as Cross-chain Interoperability Service Providers (CSPs) by  
providing required services to support cross-chain \dApps, 
such as compiling \HSL programs into transaction dependency 
graphs and speaking the \UIP protocol. 
This vision is indeed strengthened by the practical architectures of 
production blockchains, where all peer-to-peer nodes evolve 
into a hierarchy of stakeholders and a number of organizations 
operate (not necessarily own) most of the mining power for Proof-of-Work blockchains 
or/and stakes for Proof-of-Stake blockchains (whether    
such a hierarchical architecture undermines decentralization is debatable, and is out 
of the scope of \sys). Those organizations are perfectly qualified to operate as CSPs since they 
have good reachability to multiple blockchains 
and maintain sufficient token liquidity to support insurance staking, contract 
invocation, and token transfers that are required in a wide range of cross-chain \dApps. 

CSPs (\VENs) could be found via a community-driven 
\emph{directory} (similar to the Tor's relay directories~\cite{tor,tor-da}),
which we envision to be an informal list of CSPs. 
Each CSP has its own operation models, including the set of 
blockchains it has reachability, service fees charged for correct \dApp 
executions, and insurance plans to compensate CSP-introduced \dApp failures. 
Developers have full autonomy to select CSPs based on their 
\dApp requirements. Since all \dApp execution results are 
publicly verifiable, it is possible to build a CSP reputation 
system to provide a valuable metric for CSP selection.
CSPs thus misbehave at their own risk. 

Because a CSP may wish to limit its staking funds at risk in the \ISC,
a \dApp may be too large for any single CSP.
Alternatively, a \dApp may span a set of blockchains 
such that no single CSP has the reachability to 
all of them. In such cases, a cross-chain \dApp could be 
co-executed by a collection of \VENs. By design,  
\sys allows multi-\VEN executions since the 
\UIP protocol is generic to include more parties. 

We envision the industrial impact by \sys is the birth of a CSP-formed \emph{liquidity network} 
interconnected by the \UIP protocol, powering a wide range of cross-chain \dApps. 

\subsection{Complete Atomicity for \dApps}\label{discussion:atomicity}
In the context of cross-chain applications, \dApps should be 
treated as first-class citizens because the success or failure of any individual 
transaction cannot fully decide the state of \dApps. \sys 
follows this design philosophy by providing security 
guarantees at the granularity of \dApps. However, the current 
version of \sys is not fully \dApp-atomic since \UIP is unable to 
revert any state update to smart contracts if \dApps were eventually terminated prematurely. 
We recognize this as a fundamental challenge due to the finality 
guarantee of blockchains. 

To deliver full \dApp-atomicity on \sys, we 
propose the concept of \emph{stateless smart contract} where 
contracts are able to load state from blockchains before execution. 
As a result, even if the state persistent on block $\mb_n$ 
for a smart contact $\mathcal{C}$ eventually becomes dirty 
due to \dApp failure, subsequent \dApps can still load clean 
state for the contract $\mathcal{C}$ from a block (prior to $\mb_n$) agreed by 
all parties. Although this design imposes additional 
requirements on underlying blockchains, it is practical and deliverable 
using ``layer-two'' protocols where smart contract executions could 
be decoupled from the consensus layer, for instance, via the usage of 
Trusted Execution Environment (\eg Intel SGX~\cite{costan16intel} 
and Keystone~\cite{keystone}). 

\subsection{Privacy-Preserving Blockchains}\label{discussion:privacy}
The primary challenge of supporting privacy-preserving blockchains on \sys 
is the lack of a generic abstraction for those systems. 
In particular, various designs have been  
proposed to enhance blockchain privacy, such as encrypting 
blockchain state~\cite{ekiden}, obfuscating and mixing transactions via cryptography   
signature~\cite{cryptonote}. As a result, none of those blockchains 
can be abstracted as generic programmable state machines.
Therefore, our approach towards interoperating privacy-preserving blockchains  
will be application specific, such as relying on fast 
zero-knowledge proofs~\cite{bulletproofs} to allow \dApps to certify 
the state extracted from those blockchains. 

\section{Related Work}\label{sec:related}
Blockchain interoperability is often considered as one of the 
prerequisites for the massive adoption of blockchains. 
The recent academic proposals have mostly focused on moving 
tokens between two blockchains via trustless exchange protocol, including  
side-chains~\cite{chain_inter_A,pos_side,pow_side},
atomic cross-chain swaps~\cite{chain_inter_B,accs}, and 
cryptocurrency-backed assets~\cite{chain_inter_C}. 
However, programmability, \ie smart contracting across heterogeneous blockchains, 
is largely ignored in those protocols. 

In industry, Cosmos~\cite{cosmos} and Polkadot~\cite{polkadot}
are two notable projects that advocate blockchain interoperability. 
They share the similar spirit: each of them has a consensus engine to 
build blockchains (\ie Tendermint~\cite{tendermint} for Cosmos and 
Substrate~\cite{substrate} for Polkadot), and a \emph{mainchain} (\ie the Hub in 
Cosmos and RelayChain for Polkadot) to bridge individual blockchains. 
Although we do share the similar vision of ``an Internet of blockchains'', we also notice 
two notable differences between them and \sys. 
First and foremost, the cross-chain layer of Cosmos, powered by 
its Inter-blockchain Communication Protocol (IBC)~\cite{cosmos_isc}, 
mainly focuses on preliminary network-level communications. In contrast, 
\sys proposes a complete stack of designs with a unified 
programming framework for writing cross-chain \dApps and a 
provably secure cryptography protocol to execute \dApps. 
Further, at the time of writing, the most 
recent development of Cosmos and industry adoption are heading towards 
\emph{homogeneity} where only Tendermint-powered blockchains are 
interoperable~\cite{cosmos_whitepaper}. This is in fundamental contrast with \sys where 
the blockchain heterogeneity is a first-class design requirement. 
Polkadot proceeds relatively slower than Cosmos: 
Substrate is still in early stage~\cite{substrate}. 

Existing blockchain platforms such as Ethereum~\cite{ethereum-yellowpaper}
and Nebulas~\cite{nebulas} allow developers to write contracts 
using new languages such as Solidity~\cite{solidity} and Vyper~\cite{vyper}
or a tailored version of the existing languages such as Go, Javascript, and C++.
Facebook recently released Move~\cite{movelang}, 
a programming language in their blockchain platform Libra, 
which adopts the move semantics of Rust and C++ to prohibit copying and implicitly discarding coins and allow only move of the coins.
To unify these heterogeneous programming languages, we propose \HSL that 
has a multi-lang front end to parse those contacts and convert their types to unified types.
Although there exist domain-specific languages in a variety of
security-related fields that have a well-established corpus of low level
algorithms, such as secure overlay networks~\cite{mace,networklang}, network intrusions~\cite{chimera,Sommer:2014:HAE:2663716.2663735,Vallentin:2016:VUP:2930611.2930634}, and 
enterprise systems~\cite{aiql,saql},
these languages are explicitly designed to solve their domain-specific problems, 
and cannot meet the needs of the unified programming framework for writing cross-chain \dApps.


\section{Security Theorems}\label{sec:appendix:the_prof}
In this section, we present the main security theorems for our cryptography protocol \UIP, 
and rigorously prove them using the UC-framework~\cite{UC}. 

\subsection{Ideal Functionality \fbip}\label{sec:proof:ideal}
We first present the cryptography abstraction of the \UIP  
in form of an \emph{ideal functionality} \fbip. The ideal functionality articulates 
the correctness and security properties that \sys wishes to attain by 
assuming a trusted entity. The detailed description of \fbip is given in Figure~\ref{fig:fbip}. 
Below we provide additional explanations. 

\begin{figure*}
	\small
	\fbox {
		\parbox{\textwidth}{
			\vspace*{-0.15in}
			\begin{multicols}{2}
			\begin{enumerate}[label=\textbf{\scriptsize \arabic*}, leftmargin=*,itemsep=0.25ex]
				\item \Init $\storage := \emptyset$
				\item \textbf{Upon Receive} \secr($\rgt, \contract, \mathp_a, \mathp_z$):
				\item \quad generate the session ID \sid $\leftarrow \{0, 1\}^\lambda$ and keys for both parties  
				\item \quad send $\cert([\sid, \rgt, \contract]; \sigz, \siga)$ to both parties 
				\item \quad halt until both parties deposit sufficient fund, denoted as \emph{\stake} 
				\item \quad start a blockchain monitoring daemon for this session 
				\item \quad set an expiration timer \timer for executing the \contract term 
				\item \quad \textbf{for} $\mt \in \rgt$ \bco initialize the annotations for $\mt$ 
				\item \quad update $\storage[\sid] := \{ \rgt, \contract, \emph{\stake}, \timer \}$ 
				\vspace*{0.5ex}
				
					\item \textbf{Upon Receive} \rti($\mt, \sid, \mathp$):
				\item \quad $(\rgt, \_, \_, \_) := \storage[\sid]$; abort if not found 
				\item \quad assert $\mathp$ is $\mathp_z$ 
				\item \quad assert $\mt$ is eligible to be opened according the state of \gt 
				\item \quad update $\mt.\status := \init$ 
				\item \quad compute $\atte_\mt^i := \cert([\mt, \init, \sid]; \sigz)$  
				\item \quad send $\atte_\mt^i$ to both $\{\mathp_a, \mathp_z\}$ 
				to inform action 	
				
				\item \textbf{Upon Receive} \rtid($\mt, \sid, \mathp$) 
				\item \quad $(\rgt, \_, \_, \_) := \storage[\sid]$; abort if not found 
				\item \quad assert $\mathp = \mt.\from$ and $\mt.\status = \init$  
				\item \quad compute the on-chain transaction $\wtt$ for $\mt$ 
				\item \quad update $\mt.\status := \initd$ and $\mt.\trans := \wtt$
				\item \quad compute $\atte_\mt^{\id} := \cert([\wtt, \initd, \sid, \mt]; \sigpt)$  
				\item \quad send $\atte_\mt^{\id}$ to both $\{\mathp_a, \mathp_z\}$ 
				to inform action 
				
				\item \textbf{Upon Receive} \rto($\mt, \sid, \wtt, \mathp$):
				\item \quad $(\rgt, \_, \_, \_) := \storage[\sid]$; abort if not found 
				\item \quad assert $\mathp = \mt.\textsf{to}$, $\mt.\status = \initd$ and $\mt.\trans = \wtt$
				\item \quad update $\mt.\status = \open$ and get $\tsopen := \now()$ 
				\item \quad compute $\atte_\mt^{o} := \cert([\wtt, \open, \tsopen, \sid, \mt]; \sigpt)$ 
				\item \quad send $\atte_\mt^{o}$ to both $\{\mathp_a, \mathp_z\}$ 
				to inform action
				
				\item \textbf{Upon Receive} \rtod($\mt, \sid, \wtt, \mathp, \tsopen$):
				\item \quad $(\rgt, \_, \_, \_) := \storage[\sid]$; abort if not found 
				\item \quad assert $\mathp = \mt.\from$, $\mt.\status = \open$ and $\mt.\trans = \wtt$
				\item \quad assert $\tsopen$ is within the error boundary with $\now()$ 
				\item \quad update $\mt.\status = \opend$ and get $\mt.\tsopen := \tsopen$ 
				\item \quad post $\wtt$ on \fbc for on-chain execution  
				\item \quad compute $\atte_\mt^{\od} := \cert([\wtt, \open, \tsopen, \sid, \mt); \siga, \sigz)$
				\item \quad send $\atte_\mt^{\od}$ to both $\{\mathp_a, \mathp_z\}$ 
				to inform action 
				
				\item \textbf{Upon Receive} \rtc($\mt, \sid, \wtt, \tsclose$): 
				\item \quad $(\rgt, \_, \_, \_) := \storage[\sid]$; abort if not found 
				\item \quad assert $\mt.\status = \opend$  and $\mt.\trans = \wtt$ 
				\item \quad query the ledger of \fbc for $\wtt$'s status 
				\item \quad abort if $T$ is not finalized on \fbc 
				\item \quad assert $\tsclose$ is within the error boundary with current time $\now()$ 
				\item \quad update $\mt.\status := \close$ and $\mt.\tsclose := \tsclose$
				\item \quad compute $\atte_\mt^c := \cert([T, \close, \tsclose, \sid, \mt); \siga, \sigz)$  
				\item \quad send $\atte_\mt^c$ to both $\{\mathp_a, \mathp_z\}$ 
				to inform action 

				\item \textbf{Upon Receive} \tmr($\sid, \mathp \in (\mathp_a, \mathp_z)$) \public: 
				\item \quad $(\mathcal{G}_T, \contract, \emph{\stake}, \timer) := \storage[\sid]$; 
				abort if not found 
				\item \quad abort if \timer has not expired 
				\item \quad \hlgray{\# The following is the arbitration logic 
					specified by \emph{\textsf{contract}}}
				\item \quad initialize a map \resp to record which party to blame 
				\item \quad compute eligible transactions set $\ms$ given current state of \gt 
				\item \quad \textbf{for} $\mt \in \ms$ \bco 
				\item \quad \quad \textbf{if} $\mt.\status = \unknown$ \bco 
				update $\resp[\mt] := \mathp_z$ 
				\item \quad \quad \textbf{elif} $\mt.\status = \init$ \bco 
				update $\resp[\mt] := \mathp_a$ 
				\item \quad \quad \textbf{elif} $\mt.\status = \initd$ \bco 
				update $\resp[\mt] := \mt.\textsf{to}$ 
				\item \quad \quad \textbf{elif} $\mt.\status = \open$ and $\mt.\status = \opend$  \bco 
			    \item \quad \quad \quad update $\resp[\mt] := \mt.\from$ 
				\item \quad \quad \textbf{elif} $\mt.\status = \close$ \textbf{and} 
				deadline constraint fails \bco 
				\item \quad \quad \quad update $\resp[\mt] =: \mt.\from$ 
				\item \quad financially revert all \close transactions if \resp is not empty 
				\item \quad return any remaining funds in \emph{\stake} to corresponding senders  
				\item \quad remove the internal bookkeeping of \sid from \storage 

		\end{enumerate}
		\end{multicols}
		\vspace*{-0.05in}
		}
	}
	\caption{The ideal functionality \fbip.}\label{fig:fbip}
\end{figure*}

\parab{Session Setup.} Through this interface, a pair of parties 
$(\mathp_a, \mathp_z)$ (\eg a \dApp client and a \VEN) 
requests \fbip to securely execute a \dApp executable. 
They provide the executable in form of a transaction dependency graph \gt, 
as well as the correctness arbitration code \contract.
As a trusted entity, \fbip generates keys for both parties, 
allowing \fbip to sign transactions and compute 
certificates on their behalf. Both parties are required 
to stake sufficient funds, derived from the \contract, 
into \fbip. \fbip annotates each transaction wrapper $\mt$ in \gt
with its status (initialized to be \unknown), its open/close 
timestamps (initialized to 0s), and its on-chain counterpart $\wtt$ 
(initialized to be empty). To accurately match \fbip with the real-world protocol \pbip, 
in Figure~\ref{fig:fbip}, we assume that $\mathp_a$ is 
the \dApp client and $\mathp_z$ is the \VEN. 

Since \fbip does not impose any special requirements 
on the underlying blockchains, we model the ideal-world 
blockchain as an ideal functionality \fbc that supports 
two simple interfaces: \first public ledger query and 
\second state transition triggered by transactions 
(where \fbip imposes no constraint on both the 
ledger format and the consensus logic of state transitions). 

\parab{Transaction State Updates.} 
\fbip defines a set of interfaces to accept external calls for updating transaction state.  
In each interface, \fbip performs necessary correctness check to guarantee that 
the state promotion is legitimate. In all interfaces, \fbip 
computes an attestation for the corresponding transaction state, 
and sends it to both parties to formally notify the actions taken by \fbip.

\parab{Financial Term Execution.}  
Upon the expiration of \timer, both parties can invoke the 
\textsf{TermExecution} interface to trigger the contract code execution. 
The arbitration logic is also derived from decision tree mentioned in \cref{fig:decision}.
However, \fbip decides the final state of each transaction merely using its 
internal state due to the assumed trustiness. 

\parab{Verbose Definition of \fbip.} 
We \emph{intentionally} define \fbip verbosely
(that is, sending many signed messages)
in order to accurately match \fbip to
the real world protocol \pbip. For instance, in the \textsf{SessionCreate} interface, \fbip certifies 
$(\rgt, \contract, \sid)$ on behalf of both parties to 
simulate the result of a successful handshake between 
two parties in the real world. Another example is that
the attestations generated in those state update interfaces 
are not essential to ensure correctness due to the assumed trustiness of \fbip. However, 
\fbip still publishes attestations to emulate the 
\emph{side effects} of \pbip in the real world. As we shall see below, 
such emulation is crucial to prove that \fbip UC-realizes \pbip. 

\parab{Correctness and Security Properties of \fbip.}
With the assumed trustiness, it is not hard to see that 
\fbip offers the following correctness and security properties. 
First, after the pre-agreed timeout, the execution either finishes 
correctly with all precondition and deadline rules satisfied, or 
the execution fails and is financially reverted. Second, regardless 
of the stage at which the execution fails, \fbip holds the misbehaved 
parties accountable for the failure. Third, if \fbc is modeled 
with bounded transaction finality latency, \Op is 
guaranteed to finish correctly if both parties are honest. 
Finally, \fbip, by design, makes the \contract public.  
This is because in the real world protocol \pbip, 
the status of execution is public both on the \ISC and the \NSB. 
We leave the support for privacy-preserving blockchains on \sys to future work. 

\subsection{Main Security Theorems}
In this section, we claim the main security theorem of \sys. 
The correctness of Theorem~\ref{the:bip} guarantees 
that \pbip achieves same security properties as \fbip. 

\begin{theorem}\label{the:bip}
	Assuming that the distributed consensus algorithms 
	used by relevant BNs are provably secure, the hash 
	function is pre-image resistant, and the digital 
	signature is EU-CMA secure (\ie existentially unforgeable 
	under a chosen message attack), our decentralized 
	protocol \pbip securely UC-realizes the ideal functionality 
	\fbip against a malicious adversary in the passive corruption model. 
\end{theorem}

We further consider a variant of \pbip, referred to 
as \hpbip, that requires \pven and \pdapp to \emph{only use}  
\pnsb as their communication medium.

\begin{theorem}\label{lem:bip}
	With the same assumption of Theorem~\ref{the:bip}, 
	the \UIP protocol variant \hpbip securely UC-realizes the ideal functionality 
	\fbip against a malicious adversary in the Byzantine corruption model. 
\end{theorem}

\subsection{Proof Overview}\label{sec:proof_overview}
We now the prove our main theorems. We start with \cref{the:bip}.
In the UC framework~\cite{UC}, the model of \pbip  
execution is defined as a system of machines $(\me, \ma, \pi_1, ..., \pi_n)$ where $\me$ is 
called the \emph{environment}, $\ma$ is the (real-world) 
adversary, and $(\pi_1, ..., \pi_n)$ are participants 
(referred to as \emph{parties}) of \pbip where each 
party may execute different parts of \pbip. 
Intuitively, the environment $\me$ represents the 
\emph{external} system that contains other protocols, 
including ones that provide inputs to, and obtain outputs from, \pbip. 
The adversary $\ma$ represents adversarial activity 
against the protocol execution, such as controlling communication 
channels and sending \emph{corruption} messages to parties. 
$\me$ and $\ma$ can communicate freely. The \emph{passive} 
corruption model (used by Theorem~\ref{the:bip}) enables the 
adversary to observe the complete internal state of the corrupted 
party whereas the corrupted party is still protocol compliant, 
\ie the party executes instruction as desired. 
\S~\ref{sec:app:corrupt} discusses the \emph{Byzantine} corruption model, 
where the adversary takes complete control of the corrupted party. 

To prove that \pbip UC-realizes the ideal functionality \fbip, 
we need to prove that \pbip \emph{UC-emulates} 
\idealbip, which is the \emph{ideal protocol} (defined below) of our 
ideal functionality \fbip. That is, for any adversary $\ma$, 
there exists an adversary (often called simulator) $\ms$ 
such that $\me$ cannot distinguish between the 
ideal world, featured by (\idealbip, $\ms$), and the 
real world, featured by (\pbip, $\ma$). Mathematically, 
on any input, the probability that $\me$ outputs 
$\overrightarrow{1}$ after interacting with (\pbip, $\ma$) in the real world 
differs by at most a negligible amount 
from the probability that $\me$ outputs 
$\overrightarrow{1}$ after interacting with 
(\idealbip, $\ms$) in the ideal world.

The ideal protocol \idealbip is a wrapper around \fbip by a 
set of dummy parties that have the same interfaces 
as the parties of \pbip in the real world. As a result, $\me$ is 
able to interact with \idealbip in the ideal world the same way 
it interacts with \pbip in the real world. These dummy 
parties simply pass received input from $\me$ to \fbip 
and relay output of \fbip to $\me$, without implementing 
any additional logic. \fbip controls all keys of these dummy 
parties. For the sake of clear presentation, we abstract the 
real-world participants of \pbip as five 
parties \mbox{\{\pven, \pdapp, \pisc, \pnsb, \pybc\}}. The corresponding 
dummy party of \pven in the ideal world is denoted as \ipven. 
This annotation applies for other parties.  

Based on~\cite{UC}, to prove that \pbip UC-emulates 
\idealbip for any adversaries, it is sufficient to construct 
a simulator $\ms$ just for the \emph{dummy adversary} $\ma$ that 
simply relays messages between $\me$ and the parties running \pbip. 
The high-level process of the proof is that the 
simulator $\ms$ observes the \emph{side effects} of 
\pbip in the real world, such as attestation publication on the 
\NSB and contract invocation of the \ISC, 
and then accurately emulates these effects in the ideal world, with the help from \fbip. 
As a result, $\me$ cannot distinguish the ideal and real worlds. 

\subsection{Construction of the Ideal Simulator $\ms$}\label{sec:sim_cont}
Next, we detail the construction of $\ms$ by specifying what actions 
$\ms$ should take upon observing instructions from $\me$. As a distinguisher, 
$\me$ sends the same instructions to the ideal world dummy parities 
as those sent to the real world parties. 

\begin{itemize}[leftmargin=*]
	\item Upon $\me$ gives an instruction to start 
	an inter-BN session between \ipdapp and \ipven,  
	$\ms$ emulates the \gt and \contract setup 
	(\cf \S~\ref{sec:proof_hybrid}) and 
	constructs a \textsf{SessionCreate} call to 
	\fbip with parameter (\gt, \contract, \ipdapp, \ipven). 
	
	\item Upon $\me$ instructs \ipven to send an initialization request 
	for a transaction intent $\mt$,  $\ms$ extracts $\mt$ and $\sid$ 
	from the instruction of $\me$, and constructs a \textsf{ReqTransInit} 
	call to \fbip with parameter $(\mt, \sid, \ripven)$. 
	Other instructions in the same category are handled similarly by 
	$\ms$. In particular, for instruction to \xsditedb, $\ms$ 
	calls \textsf{ReqTransInited} of \fbip; for instructions to \xrvitedb, 
	$\ms$ calls \textsf{ReqTransOpen} of \fbip; for 
	instructions to \xoptb, $\ms$ calls \textsf{ReqTransOpened} 
	of \fbip; for instructions to \xcltb, $\ms$ calls \textsf{ReqTransClose} of \fbip. 
	$\ms$ ignores instructions to \xopdtb and \xcldtb. 
	$\ms$ may also extract the $\wtt$ from the instruction, which 
	is used by some interfaces of \fbip to ensure the association 
	between $\mt$ and $\wtt$.
 	
	\item Due to the asymmetry of interfaces defined by \ipdapp and \ipven, 
	$\ms$ acts slightly differently when observing instructions sent to \ipven. 
	In particular, for instructions to \xinittb, $\ms$ calls 
	\textsf{ReqTransInited} of \fbip; for instructions to \xitdtb, 
	$\ms$ calls \textsf{ReqTransOpen} of \fbip; 
	for instructions to \xoptb, $\ms$ calls \textsf{ReqTransOpened} of \fbip. 
	The rest handlings are the same as those of \ipven. 
	
	\item Upon $\me$ instructs \ipven to invoke the smart 
	contract, $\ms$ locally executes the \contract and 
	the instructs \fbip to published the updated \contract to \ipisc. 
\end{itemize}

\subsection{Indistinguishability of Real and Ideal Worlds}\label{sec:proof_hybrid}
To prove indistinguishability of the real and ideal 
worlds from the perspective of $\me$, we will go 
through a sequence of \emph{hybrid arguments}, 
where each argument is a hybrid construction 
of \fbip, a subset of dummy parties of \idealbip, 
and a subset of real-world parties of \pbip, except that the 
first argument that is \pbip without any ideal parties 
and the last argument is \idealbip without any real 
world parties. We prove that $\me$ cannot 
distinguish any two consecutive hybrid arguments. 
Then based on the transitivity of protocol 
emulation~\cite{UC}, we prove that the first 
argument (\ie \pbip) UC-emulates the last argument (\ie \idealbip). 

\parab{Real World.} We start with the real world \pbip 
with a dummy adversary that simply passes messages to and from $\me$.

\parab{Hybrid $\mathbf{A_1}$.} Hybrid $A_1$ is the same as 
the real world, except that the (\pven, \pdapp) pair is replaced by 
the dummy (\ipven, \ipdapp) pair. Upon observing an instruction 
from $\me$ to execute some \dApp executables \gt, $\ms$ calls the \textsf{CreateContract} 
interface of \pisc (living in the Hybrid $A_1$) to obtain the contract code \contract. 
Upon \contract is received, $\ms$ calls the \textsf{SessionCreate} interface of \fbip with 
parameter (\gt, \contract, \ipven, \ipdapp), which will 
output a certificate to both dummy parties to emulate the handshake 
result between \pven and \pdapp in the real world. $\ms$ also  
deploys \contract on \pnsb or \pybc in the Hybrid $A_1$. 
Finally, $\ms$ stakes required funds into \fbip to unblock 
its execution. 

Upon observing an instruction from $\me$ (sent to either dummy parties) 
to execute a transaction in \gt, 
based on its construction in \S~\ref{sec:sim_cont}, 
$\ms$ has enough information to construct a call to \fbip 
with a proper interface and parameters. If the call generates a  
certificate $\atte$, $\ms$ retrieves $\atte$ to emulate the 
PoAs staking in the real world. In particular, 
if in the real world, \pven (and \pdapp) publishes a certificate
on \pnsb after receiving the same instruction 
from $\me$, then $\ms$ publishes the corresponding 
certificate on \pnsb in the Hybrid $A_1$ as well. 
Otherwise, $\ms$ skip the publishing. 
Later, $\ms$ retrieves (and stores) the Merkle proof from 
\pnsb, and then instructs \fbip to output the proof to the dummy party 
which, from the point view of $\me$, should be the publisher of \atte. 

Upon observing an instruction from $\me$ 
(to either dummy party) to invoke the smart contract, 
$\ms$ uses its saved certificates or Merkle proofs to invoke 
\pisc in the Hybrid $A_1$ accordingly. 

Note that in the real world, the execution of \gt is automatic 
in the sense that \gt can continuously proceed even without 
additional instructions from $\me$ after successful session setup. 
In the Hybrid $A_1$, although \pven and \pdapp are replaced by dummy parties, 
$\ms$, with fully knowledge of \gt, is still able to drive 
the execution of \gt so that from $\me$'s perspective, 
\gt is executed automatically. Further, since \pisc 
still lives in the Hybrid $A_1$, $\ms$ should not trigger 
the \textsf{TermExecution} interface of \fbip to avoid 
double execution on the same contract terms. $\ms$ can 
still reclaim its funds staked in \fbip via ``backdoor'' channels 
since $\ms$ and \fbip are allowed to communicate freely 
under the UC framework. 

\parab{Fact 1.} \emph{With the aforementioned construction of $\ms$ 
	and \fbip, it is immediately clear that the outputs of both 
	dummy parties in the Hybrid $A_1$ are exactly the 
	same as the outputs of the corresponding actual parties in 
	the real world, and all side effects in the real world 
	are accurately emulated by $\ms$ in the Hybrid $A_1$. 
	Thus, $\me$ cannot distinguish with the real world and the Hybrid $A_1$.}

\parab{Hybrid $\mathbf{A_2}$.} Hybrid $A_2$ is the same as the Hybrid 
$A_1$, expect that \pisc is further replaced by the dummy \ipisc. 
As a result, $\ms$ is required to resume the responsibility of \pisc 
in the Hybrid $A_2$. In particular, when observing an instruction to 
execute a \gt, $\ms$ computes the arbitration code \contract, and then 
instructs \fbip to publish the \contract on \ipisc, 
which is observable by $\me$. 
For any instruction to invoke \contract, $\ms$ locally 
executes \contract with the input and then publishes 
the updated \contract to \ipisc via \fbip. Finally, upon 
the predefined contract timeout, $\ms$ calls 
the \textsf{TermExecution} interface of \fbip 
with parameter (\sid, \ipven) or (\sid, \ipdapp) to 
execute the \contract, which emulates the 
arbitration performed by \pisc in the Hybrid $A_1$.  

It is immediately clear that with the help of 
$\ms$ and \fbip, the output of the dummy 
\ipisc and all  effects in the Hybrid $A_2$ are exactly 
the same as those in the Hybrid $A_1$. Thus, 
$\me$ cannot distinguish these two worlds. 

\parab{Hybrid $\mathbf{A_3}$.} Hybrid $A_3$ is 
the same as the Hybrid $A_2$, expect that \pnsb is 
further replaced by the dummy \ipnsb. Since the 
structure of \pnsb and messages sent to \pnsb are public, 
simulating its functionality by $\ms$ is trivial. Therefore, 
Hybrid $A_3$ is identically distributed as Hybrid $A_2$ from the view of $\me$. 

\parab{Hybrid $\mathbf{A_4}$, \ie the ideal world.}  
Hybrid $A_4$ is the same as the Hybrid $A_3$, expect 
that \pybc (the last real-world party) is further replaced 
by the dummy \ipybc. Thus, the Hybrid $A_4$ is 
essentially \idealbip. Since the functionality of 
\pybc is a strict subset of that of \pnsb, simulating \pybc by 
$\ms$ is straightforward. Therefore, \idealbip is 
indistinguishable with the Hybrid $A_3$ from $\me$'s perspective. 

Then given the transitivity of protocol emulation, 
we show that \pbip UC-emulates \idealbip, and therefore prove 
that \pbip UC-realizes \fbip. Throughout the simulation, 
we maintain a key invariant: $\ms$ and \fbip together 
can always accurately simulate the desired outputs 
and side effects on all (dummy and real) parties 
in all Hybrid worlds. 
Thus, from $\me$'s view, the indistinguishability 
between the real and ideal worlds naturally follows. 

\subsection{Byzantine Corruption Model}\label{sec:app:corrupt}
Theorem~\ref{the:bip} considers the passive 
corruption model. In this section, we discuss the more 
general Byzantine corruption model for \pven and \pdapp 
(by assumption of this paper, blockchains and 
smart contracts are trusted for correctness). 
Previously, we construct $\ms$ and \fbip accurately to match the \emph{desired} 
execution of \pbip. However, if one party is Byzantinely 
corrupted, the party behaves arbitrarily. As a result, a Byzantine-corrupted party 
may send conflicting messages to off-chain channels and \pnsb. 
Note that for any transaction state, \pbip always processes 
the first received attestation (either a certificate from channels 
or Merkle proof from the \pnsb) and effectively ignores the other one.  
The adversary could then inject message inconsistency to make the protocol 
execution favors one type of attestations over the other. This makes it impossible 
for $\ms$ to always accurately emulate its behaviors, 
resulting in difference between the ideal world and the real world from $\me$'s view. 

To incorporate the Byzantine corruption model into our 
security analysis, we consider a variant of \pbip, referred to 
as \hpbip, that requires \pven and \pdapp to \emph{only use}  
\pnsb as the communication medium. Thus, the full granularity of protocol 
execution is guaranteed to be public and unique, allowing $\ms$ to 
emulate whatever actions a (corrupted) part may take in the real world. 
Therefore, it is not hard to conclude the \cref{lem:bip}.

\section{Conclusion}
In this paper, we presented \sys, the first platform 
that offers interoperability and programmability across heterogeneous blockchains. 
\sys is powered by two innovative designs: \HSL, a programming 
framework for writing cross-chain \dApps by unifying smart contracts written 
in different languages, and \UIP, the universal  
blockchain interoperability protocol designed to securely realize 
the complex operations defined in these \dApps on blockchains. 
We implemented a \sys prototype in approximately 35,000 lines of code to demonstrate 
its practicality, and ran experiments on the prototype to report the end-to-end execution 
latency for \dApps, as well as the aggregated platform throughput.

\section{Acknowledgments}
We thank the anonymous reviewers for their valuable feedback.
We thank Harmony Protocol for their discussion on cross-shard transactions. 
This material is based upon work partially supported by 
NSF under Contract Nos. CNS-1717313 and TWC-1518899, and by 
National Key Research and Development Program of China under grant No. 2018YFB0803605 
and NSFC under grant No. 61702045. Correspondence authors are Zhuotao Liu and Haoyu Wang. 

\balance
\bibliographystyle{acm}
\bibliography{paper} 

\clearpage
\appendix
\section{Appendix}\label{sec:appendix}
\newcommand{\epoch}{\textsf{Epoch}\xspace}
\newcommand{\rtx}{\textsf{R}_\textsf{tx}\xspace}
\newcommand{\rstate}{\textsf{R}_\textsf{state}\xspace}
\newcommand{\stroot}{\emph{\textsf{StateRoot}}\xspace}
\newcommand{\txroot}{\emph{\textsf{TxRoot}}\xspace}
\newcommand{\chainId}{\emph{\textsf{ChainID}}\xspace}
\begin{figure*}
	\small
	\fbox {
		\parbox{\textwidth}{
			\vspace*{-0.15in}
			\begin{multicols}{2}
				\begin{enumerate}[label=\textbf{\scriptsize \arabic*}, leftmargin=*,itemsep=0.25ex]
					\item \Init $\storage := \emptyset$;  $\epoch := 0$; $\ledger := []$
					
					\item \Daemon $\xdct()$ \ord:   
					\item \quad \textbf{continue} if the current \epoch is not expired 
					\item \quad $(\txp, \acp, \stp) := \storage[\epoch]$
					\item \quad initialize a block $\mathcal{B}$ with the format 
					shown in Figure~\ref{fig:nsb_block}
					\item \quad \textbf{for} \textsf{\emph{pool}} $\in$ (\txp, \acp, \stp) \bco 
					\item \quad \quad construct a (sorted) Merkle tree with \emph{selected} 
					items in \textsf{\emph{pool}} 
					\item \quad \quad populate $\mathcal{B}$ with the Merkle tree (\txmt, \acmt, or \stmt) 
					\item \quad \quad remove these selected items from \textsf{\emph{pool}}
					\item \quad update $\ledger.\textsf{append}(\mathcal{B})$ and execute 
					trans. captured under \txmt
					\item \quad start a new epoch $\epoch := \epoch + 1$ 
					\item \quad initialize $\storage[\epoch] := [\txp, \acp, \stp]$ 
					
					\item \Daemon \cc($\wtt$, [\chainId, \stroot, \txroot]): 
					\item \quad $(\_, \_, \stp) := \storage[\epoch]$; abort if not found
					\item \quad update $\stp.\add(\wtt)$ 
					\item 
					\noindent \adjustbox{bgcolor=gray!20,minipage=[t]{\linewidth}}{
						\quad \# \emph{Protocol of an individual honest peer $\mv$ to ensure correctness} 
						\item \quad \Init $\mv.\stp := []$ 
						\item \quad \Daemon \probe($\wtt, \ms=\{\sig^{\textsf{peer}}, ...\}$) 
						\item \quad \quad abort if $\wtt$ is already in $\mv.\stp$ 
						\item \quad \quad 
						abort if $\ms$ contains more than $\mk$ distinguished signatures 
						\item \quad \quad abort if $\wtt$ is not finalized on its destination blockchain (\chainId)
						\item \quad \quad abort if the reported \stroot and \txroot are not authentic
						\item \quad \quad update $\mv.\stp.\add(\wtt)$ 
						\item \quad \quad update $\ms.\add(\cert([\wtt, [\chainId, \stroot, \txroot]]; \sig^\mv))$ 
						\item \quad \quad multicast ($\wtt, \ms$) to other peers of the \NSB 
					}
					
					\item \Daemon \cw(\{\pbc, ...\}): 
					\hspace*{\fill} \textcolor{olive}{\emph{Proactive Streaming Version}}
					\item \quad proactively watch \pbc for recently finalized blocks $\{\mathcal{B}, ...\}$
					\item \quad $(\_, \_, \stp) := \storage[\epoch]$; abort if not found
					\item \quad retrieve the root $\rtx$ of \txmt and $\rstate$ of \statemt on $\mb$
					\item \quad update $\stp.\add(\rtx, \rstate)$ 
					\item 
					\noindent \adjustbox{bgcolor=gray!20,minipage=[t]{\linewidth}}{
						\quad \# \emph{Protocol of an individual honest peer $\mv$ to ensure correctness} 
						\item \quad \Init $\mv.\stp := []$ 
						\item \quad \Daemon \probe(\{\pbc, ...\}) and \probe($\mb, \ms = \{ \sig, ... \}$)
						\item \quad \quad abort if $\mb$ is not finalized on \pbc or $\mb$ is processed before
						\item \quad \quad 
							abort if $\ms$ contains more than $\mk$ distinguished signatures 
						\item \quad \quad retrieve the root $\rtx$ of \txmt and $\rstate$ of \statemt on $\mb$
						\item \quad \quad $\ms.\add(\cert([\rtx, \rstate]; \sig^\mv))$; $\mv.\stp.\add(\rtx, \rstate)$ 
						\item \quad \quad multicast ($\mb, \ms$) to other peers of the \NSB 
					}
										
					\item \textbf{Upon Receive} $\xadda(\atte)$:
					\item \quad $(\_, \acp, \_) := \storage[\epoch]$; abort if not found
					\item \quad update $\acp.\add(\atte)$ 
					\item 
					\noindent \adjustbox{bgcolor=gray!20,minipage=[t]{\linewidth}}{
						\quad \# \emph{Similar correctness protocol as in \xccb for honest peers} 
					}
					
					\item \textbf{Upon Receive} $\xexec(\wtt)$ \ord:
					\item \quad abort if $\wtt$ is not correctly constructed and signed 
					\item \quad $(\txp, \_, \_) := \storage[\epoch]$; abort if not found
					\item \quad update $\txp.\add(\wtt)$ 
					
					\item \textbf{Upon Receive} $\xbh()$ \ord:
					\item \quad return the block number of the last block on \ledger 
					
					\item \textbf{Upon Receive} $\xmkp(\textsf{key})$ \ord: 
					\item \quad find the block $\mathcal{B}$ on \ledger containing 
					\textsf{key}; abort if not found 
					\item \quad return a hash chain from \textsf{key} to the  
					Merkle root on $\mathcal{B}$ 
				\end{enumerate}
			\end{multicols}
			\vspace*{-0.05in}
		}
	}
	\caption{Detailed protocol description of \fnsb. 
		Interfaces annotated with \ord are also 
		implemented by \pbc. \hlgray{Gray background} 
		denotes the protocol of honest peers in the \NSB to 
		ensure the correctness for the corresponding 
		interface.}\label{fig:fnsb}
\end{figure*}

\subsection{Specification of \fnsb and \pbc}\label{sec:appendix:fnsb_spec}
The detailed protocol description of \fnsb and \pbc is given in \cref{fig:fnsb}. 
We model block generation and consensus in \fnsb (and \pbc) as 
a \emph{discrete clock} that proceeds in \emph{epochs}. The length 
of an epoch is not fixed to reflect the consensus. 
At the end of each epoch, a new block is packaged and added to the append-only 
public \ledger. 

The block format of \fnsb is shown in \cref{fig:nsb_block}. 
Each block packages two special Merkle trees (\ie \acmt and \stmt) 
and other Merkle trees (\eg \txmt) that are common in both \fnsb 
and \pbc. \stmt and \acmt  are constructed using the items 
in the \stp and \acp, respectively. Considering the size limit of one block, some items in 
these pools may not be included (\eg due to lower gas prices or random selections), 
and will be rolled over to the next epoch. 

\stp is constructed via the \xccb interfaces executed by all \NSB peers. 
The \xccb directly listens for transactions claimed by \sys users (\VENs and \dApp clients). 
This design avoids including irrelevant transactions 
that are not generated by any \sys sessions into the \stmt. 
An alternative design for building \stp is via the \xcwb which proactively watches all underlying 
blockchains to collected the transaction Merkle roots 
and state roots packaged in recently finalized blocks. This design 
is more cost efficient for \sys users since they do not need to explicitly claim 
these roots. Meanwhile, the structure of the \stmt is changed to the same as 
\acmt since all its stored data now become static. 

Figure~\ref{fig:fnsb} further specifies the protocol of each individual   
honest peer/miner in the \NSB to ensure the correctness of 
both interfaces. By complying with the protocol, honest peers accept any received claim 
(\ie a Merkle root or transaction claim) only after receiving a quorum of approvals for the claim. 
The protocol provably ensures the correctness of 
both interfaces, given that the number 
of Byzantine nodes in the permissioned \NSB is no greater than the security 
parameter $\mathcal{K}$~\cite{byzantine}. 

The \acp is constructed in a similar manner as the \stp. 
In Figure~\ref{fig:fnsb}, an interface annotated with \ord 
is also implemented by \pbc, although the implementation 
detail may be different; for instance \pbc may have different 
consensus process than \fnsb in the \xdctb interface.

\end{document}